\documentclass[paper]{JHEP3}

\usepackage{epsfig,multicol,bbm}
\usepackage{here}
\usepackage{amsmath}

\newcommand {\beq} {\begin{equation}}
\newcommand {\eeq} {\end{equation}}
\newcommand {\beqa}{\begin{eqnarray}}
\newcommand {\eeqa}{\end{eqnarray}}
\newcommand {\del} {\partial}

\newcommand {\Tr}{\mbox{Tr\,}}

\newcommand {\ee}{\mbox{e}}

\title{Direct test of
the gauge-gravity correspondence
for Matrix theory correlation functions}

\author{Masanori Hanada\\Department of Physics, University of Washington, 
 Seattle, WA 98195-1560, USA,\\
E-mail: \email{mhanada@u.washington.edu}}

\author{Jun Nishimura\\KEK Theory Center, High Energy Accelerator
Research Organization (KEK),\\
1-1 Oho, Tsukuba, Ibaraki 305-0801, Japan, and\\
Department of Particle and Nuclear Physics,
School of High Energy Accelerator Science,\\
Graduate University for Advanced Studies (SOKENDAI),\\
1-1 Oho, Tsukuba, Ibaraki 305-0801, Japan,\\
E-mail: \email{jnishi@post.kek.jp}}

\author{Yasuhiro Sekino\\KEK Theory Center, High Energy Accelerator
Research Organization (KEK),\\
1-1 Oho, Tsukuba, Ibaraki 305-0801, Japan,\\
E-mail: \email{sekino@post.kek.jp}}

\author{Tamiaki Yoneya\\ 
School of Graduate Studies, 
The Open University of Japan,\footnote{New address after April 1, 2010, 
moving from the Institute of Physics, University of Tokyo, Komaba.}\\
2-11 Wakaba, Mihama-ku, Chiba 261-8586, Japan,\\
E-mail: \email{tam@hep1.c.u-tokyo.ac.jp}}

\preprint{KEK-TH-1490}	

\abstract{
We study correlation functions
in (0+1)-dimensional
maximally supersymmetric U($N$) Yang-Mills theory, 
which was proposed by Banks et al.\ as a
non-perturbative definition of 11-dimensional M-theory in the infinite-momentum frame.
We perform first-principle calculations using Monte Carlo simulations,
and compare the results against
the predictions obtained previously based on 
the gauge-gravity correspondence from 10 dimensions.
After providing a self-contained review on these predictions,
we present clear evidence that the 
predictions
in the large-$N$ limit 
actually hold even
at small $N$ such as $N=2$ and $3$.
The predicted behavior
seems to continue to the far infrared regime,
which goes beyond the naive range of validity of
the 10D supergravity analysis. 
This suggests that the correlation functions also
contain important information on the M-theory limit. 
}


\begin{document} 


\section{Introduction}
It is widely recognized nowadays that gauge theory plays
an important role in understanding quantum gravity and string theory.
In this context 
maximally supersymmetric U($N$) Yang-Mills theory
in (0+1) dimension is of particular interest.
This theory was proposed as a non-perturbative formulation of 
M-theory by
Banks et al.\ (BFSS) \cite{BFSS} in an appropriate large-$N$ limit. 
It is also called, justifiably, ``Matrix theory,'' 
the term which we will use throughout the present paper 
even for finite $N$
for the sake of simplicity of nomenclature.
M-theory is a conjectural 11-dimensional theory~\cite{mtheory, mthtownsend}, 
which was suggested
to play a pivotal role in achieving a complete unification 
of perturbative superstring theories.
When the 11th dimension of M-theory is 
compactified to a circle S$^1$ or S$^1$/Z$_2$ of radius $R$,
the effective 10-dimensional theory 
one obtains at small $R$ is assumed to be
perturbative type IIA or Heterotic E$_8 \times$ E$_8$ superstring theory, 
respectively, where the radius is given as $R=g_s\ell_s$ 
in terms of the string coupling constant $g_s$ and
the string length $\ell_s$.
These relations, together with the string dualities in 10 
(or 9) dimensions, connect all the perturbative superstring theories. 
It is therefore naturally expected that deeper principles 
behind string theory will be revealed 
if one finds a concrete realization of M-theory.

The action of Matrix theory coincides
with that of the effective super Yang-Mills theory describing 
a collection of $N$ D0-branes \cite{wittenbound}
in type IIA superstring theory in 10 dimensions,
where the Yang-Mills coupling $g_{\rm YM}^{\, 2}$ is
proportional to the string coupling $g_s$.
Each D0-brane with the mass $1/g_s\ell_s$ can be regarded as a Kaluza-Klein state of the 11D graviton 
carrying one unit of momentum along the compactified 11th direction.
The proposal advocated in ref.~\cite{BFSS} was that 
Matrix theory in the large-$N$ limit with fixed $g_{\rm YM}^{\, 2}$ 
provides the exact quantum description of  M-theory 
in spatial 9-dimensional transverse directions of 
the infinite momentum frame that is boosted along the circle
in the 11th dimension.
 
Another closely related motivation 
for this (0+1)-dimensional theory is that it can alternatively be 
regarded as a regularized realization~\cite{supermembrane} 
of quantum (super)membrane theory in the light-cone gauge, 
again in the large-$N$ limit. 
 Supermembranes have also been 
regarded as basic degrees of freedom 
in M-theory, playing similar roles as strings in
10D superstring theory at least in some particular 
long-distance regime. Matrix theory can therefore be 
a natural basis for approaches to M-theory from the viewpoint of supermembranes, 
which has been quite an elusive subject for the last few decades. 

     From these two interpretations,  
the D0-branes are regarded as ``partons'' for 
both gravitons and membranes in 11 dimensions, 
and Matrix theory is supposed to describe the 
dynamics of composite systems consisting of a 
large number of D0-partons. 
Since the 't Hooft coupling constant 
 $\lambda 
=g_{{\rm YM}}^{\, 2}N$ of Matrix theory must be 
infinitely large in the M-theory interpretation,
one has to consider the strongly coupled regime of the 
Yang-Mills theory. 
Understanding the strong coupling dynamics 
is expected also to shed light on important problems 
in quantum gravity such as a microscopic 
description of the formation and evaporation of black holes 
in terms of D-branes. 
In spite of such prospects, 
it seems fair to say that 
no substantial progress has been made 
with respect to the dynamics of
Matrix theory after the end of the previous century. 
Unfortunately, M-theory itself still remains an enigma.


Although Yang-Mills theory
in $(0+1)$ dimensions 
is formally super-renormalizable 
with respect to the UV behavior,
there is very little knowledge 
on its strong coupling dynamics, which is essential 
for understanding the theory in the IR regime.
A crucial obstacle is the severe IR divergence 
that comes from loop effects of massless fields
in calculating correlation functions.
While it is expected that some mass scale
generated non-perturbatively
makes the correlation functions finite,
perturbation theory does not seem to give us useful insight 
on such a mechanism. 
In fact, by introducing mass scales through
background fields, one can perform perturbative calculations,
which nicely demonstrate
that Matrix theory can indeed describe gravitational interactions 
of D0-branes in agreement with the results from supergravity. 
A remarkable example is the derivation of the general 
relativistic three-body forces~\cite{okawayoneya}  
from the gauge theory side. This shows that 
the basic structure of the interaction of slowly moving 
D0-branes 
is indeed consistent with supergravity in the weakly coupled regime. 
However, for the purpose of 
studying correlation functions that would carry 
crucial information on the wave functions 
of D0-brane bound states, we cannot assume any 
preferred background fields. 

On the other hand, the gauge-gravity correspondence 
has been studied intensively over the past decade
as a possible powerful approach to strongly coupled gauge theories
including QCD and condensed matter systems.  
The best tested example is 
the AdS-CFT correspondence \cite{Maldacena:1997re}
between 4D ${\cal N}=4$ SU($N$) super Yang-Mills theory (SYM) and 
type IIB superstring theory on the 
special background ${\rm AdS}_5\times {\rm S}^5$. 
Both sides are characterized by exact (super) conformal  
symmetry, which greatly facilitates the analyses. 
From the viewpoint of string dynamics, this correspondence 
was motivated from the open-string/closed-string duality. 
In particular, if one takes the large-$N$ limit 
with a large but {\it fixed} 't Hooft coupling $\lambda$ on the SYM side, 
both the gravitational coupling constant $G_{10}
\sim g_s^{\, 2} \, \ell_s^{\, 8}$ 
and the curvature ($\sim (g_s N)^{-1/4} \, \ell_s^{\, -1}$) of 
the background geometry become small 
on the string theory side in the bulk.
Therefore, one can study the gauge theory in that limit
by using classical supergravity theory, which describes 
the low-energy limit of string theory. 

In view of the general arguments 
based on the 
open-string/closed-string duality, 
there seems 
nothing that restricts the gauge-gravity correspondence
to the cases with exact conformal symmetry.
In particular, 
Matrix theory is expected to be dual to type IIA superstring theory 
on the background obtained in the near-horizon limit 
of the D0-brane solution~\cite{IMSY}.
Although neither the action of Matrix theory 
nor the near-horizon background of D0-branes is
invariant under conformal transformations,
one can transform them back to its original form
by allowing the coupling constant to transform 
appropriately.
This 
``generalized" conformal symmetry \cite{Jevicki, Jevicki:1998ub}
turned out to be useful in formulating 
the gauge-gravity correspondence at the level of operators. 
Unlike the familiar AdS cases,
the dilaton and the curvature 
in the present non-conformal case are not constant in space-time
in the near-horizon limit of the D0-brane solution. 
In fact the string coupling becomes stronger towards
the origin (which corresponds to the IR region in the SYM), while 
the curvature becomes stronger towards the boundary
(which corresponds to the UV region in the SYM).
Thus there is a certain region in which we
expect the supergravity description to be valid. 
In order for this region to extend both in the UV and IR directions,
$N$ must be sent to $\infty$ with 
a fixed but large 't Hooft coupling constant.
However, the region of validity of this correspondence is slightly 
different from the BFSS large-$N$ limit that is relevant to M-theory. 
This is not so surprising since the gauge-gravity 
correspondence is based on the 
D0-brane picture in 10-dimensional space-time, and
the space-time dimension must somehow be elevated to 11 dimensions 
for the M-theory interpretation. 
Let us recall here that 
the dilaton carries information on the effective 
size of the 11th direction. 
Therefore, one may hope that 
the predictions from 10-dimensional supergravity 
contain
important information on the 11-dimensional theory 
as one effectively probes the 11th direction 
by approaching the center of the D0-brane solution. 
This is one of the main issues we would like to 
address in the present work. 
%

Using the gauge-gravity correspondence, 
two of the present authors (Y.S.\ and T.Y.)
obtained two-point correlation functions in Matrix theory 
from supergravity on the D0-brane background \cite{SY}. 
This work was based on an extension of the general prescription 
proposed by Gubser, Klebanov, Polyakov 
and independently by Witten (GKPW)~\cite{GKPW} 
for the standard AdS case, which enables us to compute correlation functions
in gauge theory by evaluating the supergravity action 
as a functional of boundary values of the fields. 
Natural candidates for operators corresponding to the supergravity modes
are given by the Matrix theory currents, 
as suggested from a perturbative analysis 
with non-trivial D0-brane backgrounds \cite{Taylor}. 
They are single-trace operators with totally symmetric 
ordering inside the trace, 
which may be viewed as analogs of the BPS-operators in the ${\cal N}=4$ SYM. 
The dictionary between these operators and 
the supergravity spectrum on the D0-brane background was given
in ref.~\cite{SY} by matching the scaling dimensions 
with respect to the generalized conformal symmetry.

For the operators corresponding to the massless 
modes, which can be described by the supergravity 
approximation, 
the two-point correlation functions
are predicted to take the general form 
\begin{equation}
 \Big\langle
{\cal O}(t) \, {\cal O}(t')
\Big\rangle
\propto
\frac{1}{|t-t'|^{2\nu+1}} \ .
\label{coordspace} 
\end{equation}
The supergravity approximation with the near-horizon geometry 
is valid in the region  
\begin{equation}
\lambda^{-1/3} \ll |t-t'|\ll
\lambda^{-1/3} N^{10/21} \ .
\label{region} 
\end{equation}
The exponent $\nu$ in (\ref{coordspace})
turned out to be
fractional numbers.  They are different
from those determined by the canonical dimensions of the operators, 
but are consistent
with the generalized conformal symmetry. 
The fact that the exponents are fractional numbers 
suggests
that confirming this result from the gauge theory side 
would 
require
a completely non-perturbative calculation of the exponents.
On dimensional grounds they must be independent 
of the Yang-Mills coupling constant, 
which has a positive mass dimension. 
It should also be emphasized on the same grounds 
that the power-law behavior itself is quite non-trivial. 
This suggests that some mechanism for decoupling the 
non-perturbative dynamics
from the mass scale of the Yang-Mills coupling constant 
is at work
for this class of operators, 
possibly owing to supersymmetry. 

From the viewpoint of the Matrix-theory 
conjecture, on the other hand, 
the power-law behavior in the IR regime is an 
important dynamical signature 
of the theory~\cite{Y}.
Namely, it is related to 
the existence of a threshold bound state (See ref.~\cite{threbound} 
and references therein.)
at each finite $N$. This is required for consistent
identification of D0-branes \cite{wittenbound} as partons 
for 11-dimensional gravitons. 
For unprotected operators, on the other hand, it is natural to expect that 
two-point correlation functions are coupled with 
the Yang-Mills dynamics of mass generation and are 
given in the form of 
exponential functions of some dimensionless combination 
of the coupling constant and the coordinate 
distances.
Indeed, such a behavior \cite{ASY} has been predicted 
for operators 
corresponding to stringy excited states 
with large angular momentum 
in the plane-wave limit of the background geometry.

In the present paper, we study the two-point correlation functions 
by performing Monte Carlo simulation of Matrix theory.
The same method~\cite{Hanada-Nishimura-Takeuchi} 
has been previously used to calculate 
the internal energy and the Wilson loop in Matrix theory
at finite temperature by the authors including 
two of us (M.H.\ and J.N.)~\cite{AHanada-Nishimura-Takeuchi,%
Hanada-Hyakutake-Nishimura-Takeuchi,HMNT}.
The results of these works are 
consistent with the 
predictions from the dual geometry in the bulk. 
For example, 
the ADM mass of the 
near extremal D0-brane solution 
gives the energy density on the gauge theory side 
as a function of temperature as\footnote{The 
power-law behavior (\ref{E-Trel})
is also related to 
the existence of the threshold bound state \cite{Smilga:2008bt} 
and hence to the power-law behavior of the correlation functions. 
In contrast,
Monte Carlo studies for
the non-maximally supersymmetric cases \cite{Hanada:2010jr}
confirmed an exponential decrease at low $T$, which was
suggested in ref.~\cite{Smilga:2008bt} by assuming the absence
of the threshold bound state.}
\begin{eqnarray}
\frac{1}{N^2} \frac{E}{\lambda^{1/3}}
=
7.4 \left( \frac{T}{\lambda^{1/3}} \right)^{14/5} \ , 
\label{E-Trel}
\end{eqnarray}
assuming that the temperature is identified with the Hawking 
temperature of the black hole. This relation is 
expected to hold under the condition 
$N^{-10/21}\ll T\lambda^{-1/3}\ll 1$,
which coincides naturally with (\ref{region}) under the 
replacement $1/T\rightarrow |t-t'|$. 
The Monte Carlo data on the gauge theory side not only confirms
this prediction but also gives 
consistently the power of the sub-leading term 
with respect to the inverse YM coupling that are 
obtained from the $\alpha ' (=\ell_s^{\, 2})$ corrections
on the gravity (string theory) 
side \cite{Hanada-Hyakutake-Nishimura-Takeuchi}. 
The agreement 
implies that the microscopic origin of 
the black-hole thermodynamics has been accounted for 
in terms of open strings attached to the D0-branes
as described by Matrix theory. 
These results 
encourage us to apply
the same method to more  
general and subtle observables
such as the correlation functions of local 
operators.
Let us recall that 
the correlation functions are calculated
on the gravity side \cite{SY} 
by extracting the response from the D0-brane background 
against perturbations consisting of various different 
components of gravitational waves 
propagating from the boundary towards the central region. 
Such observables therefore
contain more detailed information 
on the dynamics of D0-branes than the observables
that have been studied so far.

The aim of the present paper is twofold.
Firstly, we give an explicit comparison of the Monte Carlo data with 
the predictions for supergravity states \cite{SY}
and also 
with those for stringy excited states \cite{ASY}. 
This provides, for the first time, a highly nontrivial test
of the gauge-gravity correspondence 
at the level of local operators in the non-conformal case.\footnote{
See ref.~\cite{Hiller} for 
a numerical analysis based
on the discrete light-cone quantization
in the $(1+1)$-dimensional case.
} 
Secondly, we provide evidence for the possibility that
the predictions obtained from the gauge-gravity correspondence 
actually hold beyond the naive bounds 
of the applicability of supergravity approximation, 
both with respect to the size $N$ of the gauge group and 
to the range (\ref{region}). 
This 
may be taken as
an important piece of information on the 
behaviors of correlation functions in the M-theory regime. 
We hope that the present work\footnote{Part of the results
has been reported briefly in our previous publication \cite{HNSY}.}  
would be an 
impetus for revisiting
the Matrix theory conjecture from 
a new perspective.





The rest of this paper is organized as follows. 
In section 2, we review the gauge-gravity correspondence 
for Matrix theory and summarize its predictions 
for correlation functions.
We hope that 
this section 
makes the present paper 
reasonably self-contained and will serve as a useful guide
for the readers who are not familiar with 
the gauge-gravity correspondence for D0-branes and Matrix theory. 
In section 3, we describe our Monte Carlo method to calculate 
the correlation functions and 
compare the results with the predictions 
from the gauge-gravity correspondence.
Section 4 is devoted to a summary
and discussions.
%
%
In appendix \ref{sec:spectral-rep}, we discuss the physical meaning
of the predicted two-point functions for stringy
operators
by considering its spectral representation.
In appendix \ref{sec:detail-MC}, 
we explain the details of our Monte Carlo method.
In appendix \ref{sec:sign-problem}, 
we discuss 
the so-called sign problem 
that appears in Monte Carlo studies of the present system.


\section{Matrix theory and the gauge-gravity correspondence}
\label{sec:gravity_side}

In this section we first give a brief introduction
to Matrix theory, which was 
proposed as a non-perturbative
formulation of M-theory,
and discuss the D0-brane solution in 10-dimensional
supergravity, which plays a crucial role in describing
the gauge-gravity correspondence in the present case.
Then we review the predictions for correlation functions 
obtained from the gauge-gravity correspondence
to the extent that would be sufficient 
for understanding the significance of our main results
to be presented in section \ref{sec:MCresults}.

\subsection{brief review of Matrix theory}
\label{action-Mtheory}

Matrix theory is the maximally supersymmetric Yang-Mills theory
in (0+1) dimensions, and it can be considered as matrix quantum mechanics.
The action is given by
\begin{equation}
S =\int dt\ 
{\rm Tr}
\left(
\frac{1}{2g_s \ell_s} \, (D_tX_i)^2
+ \frac{1}{4g_s\ell_s^{\, 5}} \, 
[X_i,X_j]^2
+ \frac{i}{2} \, \psi_\alpha D_t\psi_\alpha -
\frac{1}{2\ell_s^{\, 2}}\, \psi_\alpha(\gamma_i)_{\alpha\beta}
[X_i,\psi_\beta]
\right) \ ,
\label{Matrixaction}
\end{equation}
where the fields $X_i\ (i=1,\cdots , 9)$ and $\psi_\alpha\ 
(\alpha=1,\cdots,16)$ are 
$N\times N$ bosonic and fermionic Hermitian matrices 
representing the collective degrees of freedom and 
their superpartners associated with 
the D0-branes. The theory has a U($N$) gauge symmetry,
and the covariant derivative $D_t$ acts as 
$D_t=\partial_t-i \, [A,\ \cdot\ ]$ with the gauge field $A$.
The $16 \times 16$ gamma matrices $\gamma_i$ 
satisfy the SO(9) Clifford algebra 
$\{ \gamma_i,\gamma_j \}= 2 \, \delta_{ij}$.
The Yang-Mills coupling $g_{\rm YM}$ is related to the string
coupling $g_s$ and the string length $\ell_s$
as $g_{\rm YM}^{\, 2}=(2\pi)^{-2}g_s \, \ell_s^{\, -3}$.
While the action (\ref{Matrixaction}) is written 
with the Lorentzian signature, 
we will 
use the Euclideanized time coordinate
in actual calculations both on the supergravity side and 
on the gauge theory side.

From the viewpoint of open string theory, 
the above action is regarded as the low-energy 
effective theory for the collection of $N$ 
D0-branes in type IIA superstring theory in 10 dimensions~\cite{wittenbound}. 
The low-energy approximation is justified 
in the non-relativistic limit, where $D_tX_i$ and higher 
derivatives are sufficiently small. 
The matrices $X_i$ describe the lowest modes of open strings 
connecting the D0-branes, and their diagonal components represent the 
positions of the D0-branes in the 9-dimensional space. 
The U(1) part, which is decoupled from the SU($N$) part,
corresponds to the center of mass motion of the D0-branes.
The SU($N$) part, on the other hand,
represents the relative motions of the D0-branes 
and their interactions through open strings. 

From the viewpoint of 11-dimensional M-theory,  
the D0-brane states 
represent the Kaluza-Klein modes 
of the graviton supermultiplet corresponding to 
the compactification 
down to 10 dimensions. 
The radius of the 11th direction is $R=g_{s} \ell_s$. 
The BFSS conjecture~\cite{BFSS} is a proposal 
that Matrix theory in the large-$N$ limit with fixed $g_s$ 
describes exactly 
the dynamics of 11-dimensional gravitons 
in the infinite momentum frame (IMF) 
boosted along the 11th direction with the longitudinal momentum $P^+=N/R$. 
The 9 spatial directions in the 10-dimensional 
interpretation now correspond to the transverse directions 
in the IMF. 
The non-relativistic approximation 
is justified in this frame since $P^+$ becomes infinitely large in the above 
limit. 
Note that the large-$N$ limit 
considered by BFSS
is different from the 't Hooft limit, in which
$g_{{\rm YM}}^{\, 2} N$ is fixed. 
Let us also note that the ordinary weak coupling limit with 
finite $N$ and small $g_s$ is another way to realize 
$P^+\rightarrow \infty$ and hence the IMF. This alternative 
is usually called 
 the ``discrete light-cone quantization''
(DLCQ).

The Matrix theory (\ref{Matrixaction})
is not scale invariant in the usual sense
since the Yang-Mills coupling constant has a mass dimension.
However, it has a generalized conformal symmetry (GCS),
which can be naturally understood from the viewpoint of 
the D0-brane dynamics \cite{Jevicki}.
As emphasized in ref.~\cite{Jevicki}, a deeper motivation 
for the GCS originates from the space-time uncertainty principle 
in string theory.\footnote{For a 
comprehensive discussion on the idea of the space-time uncertainty 
relation in string theory, see ref.~\cite{Ystu},
which also contains an extensive list of references.}
Under the GCS, the coupling
constant $g_{\rm YM}^{\, 2} \propto g_s$ 
is scaled (with the scaling dimension 3)
along with the coordinates and the fields as
\begin{equation}
t \to a^{-1}t \ , \quad 
X \to a X \ , \quad 
g_s \to a^{3}g_s \ .
\label{GCS}
\end{equation} 
Let us recall that the string coupling $g_s 
=e^{\langle \phi \rangle}$ is related to the 
vacuum expectation value ${\langle \phi \rangle}$ of the dilaton in the 
asymptotic region. 
All the other generators including the special conformal transformation 
can also be defined, and
the full conformal algebra is actually realized~\cite{Jevicki}. 
One can also see that the scaling transformation (\ref{GCS})
corresponds to the Lorentz boost in the IMF since one can combine
it with
a trivial engineering rescaling $(t\rightarrow a^{-1} t, 
X\rightarrow a^{-1} X, \ell_s
\rightarrow a^{-1}\ell_s)$
to obtain the boost transformation 
$(t\rightarrow a^{-2}t, X\rightarrow X, 
R\rightarrow a^2 R)$ with the 11D Planck 
length $\ell_P=g_s^{\, 1/3}\ell_s$ fixed. 
Thus the GCS is consistent with the 
DLCQ interpretation of Matrix theory. To test the boost symmetry
in the BFSS interpretation, on the other hand, 
we need to examine a much 
more dynamical aspect of the theory, since 
it is related to the scaling behavior in changing $N$. 
(See section \ref{sec:prediction-sugra}.)

In passing we note that the Hamiltonian of this system 
is given as 
\begin{equation}
H = -2 P^- = R \, 
{\rm Tr}
\left(
\frac{1}{2}\, \Pi^2-\frac{1}{4g_s^{\, 2} \, \ell_s^{\, 6}}
\, [X_i, X_j]^2
+\cdots \right) \ ,
\label{Ham-Pminus}
\end{equation}
where $\Pi_i$ are the canonical (matrix) momenta  corresponding to the 
matrix coordinates $X_i$. 
In the M-theory limit, $P^+P^-$ is fixed and $P^+ \propto 
N$ is sent to $\infty$. Thus we have to 
consider the spectrum of $H$ which scales as $1/N$ 
near the threshold $H\sim 0$. 
In terms of two-point correlation functions, 
this amounts to considering the far IR region
in which the time coordinate scales linearly in $N$.  

\subsection{D0-brane solution and its near-horizon limit}
\label{sec:D0-sol}
The crucial point for the gauge-gravity correspondence is that
the D0-branes, which are described by the action (\ref{Matrixaction})
from the viewpoint of open strings,
can also be described 
as a classical background in 10-dimensional supergravity
from the viewpoint of closed strings.
The solution~\cite{Horowitz:1991cd} 
corresponding to a stack of $N$ D0-branes is given 
in the string frame as
\begin{eqnarray}
&& ds^2=-h^{-1/2}dt^2+h^{1/2}
\Bigl(dr^2+r^2d\Omega_8^2 \Bigr) \ ,\nonumber \\
&&e^{\phi}=g_{s} \, h^{3/4} \ , \quad A_0=-\frac{1}{g_s}
\Bigl(h^{-1}-1 \Bigr) \ , \nonumber \\
&&h=1+{q\over r^7} \ , \qquad q=60\pi^3 (g_s N) \,\ell_s^{\, 7} \ ,
\label{orginal-D0metric}
\end{eqnarray}
where $\phi$ and $A_0$ represent the dilaton and the RR gauge field,
respectively.
String theory (or supergravity) on this background in the
near horizon limit
$h\to {q\over r^7}$
is conjectured to be equivalent to Matrix theory~\cite{IMSY,Jevicki}
under certain 
conditions explained below.
Unlike the case of ${\rm AdS}_5\times {\rm S}^5$, which
corresponds to the near-horizon limit of the D3-brane solution, 
both the dilaton and the curvature depend on the radial coordinate.
From the explicit form of the background, one can obtain 
the range of validity
\begin{equation}
(g_s N)^{1/3}N^{-{4/21}}\ll 
\frac{r}{\ell_s} 
\ll (g_s N)^{1/3} \ , 
\label{validity}
\end{equation}
in which the weakly coupled supergravity description is reliable.
The first inequality is necessary for the string coupling to be weak
$e^{\phi}\ll 1$, 
while the second one is necessary for the curvature 
to be small in the string unit.  
Taking the near-horizon condition 
$\frac{r}{\ell_s} \ll (g_s N)^{1/7}$ 
also into account, 
we find that a wide range of validity can be obtained
if we consider the 't Hooft limit $N \to \infty$ 
in the strongly coupled region, namely $g_sN \gg 1$.

Let us note here that the near-horizon limit of the 
D0-brane metric can be obtained by a Weyl transformation 
from the ${\rm AdS}_2\times {\rm S}^8$ metric as
\begin{equation}
ds^2=q^{2/7} h^{3/14} \left\{\left({2\over 5}\right)^2
{-dt^2+dz^2\over z^2}+
\Bigl( d\theta^2+\cos^2\theta d\psi^2+\sin^2\theta
d\Omega_6^2 \Bigr) \right\}  \ . 
\label{D0metric}
\end{equation}
Here the Poincar\'{e} coordinate $z$ is related to $r$ by 
\begin{equation}
z={2\over 5} \, q^{1/2}r^{-5/2} \ ,
\label{poincare-coord}
\end{equation}
and we have chosen a special parametrization of S$^8$, 
which will be useful in section \ref{sec:pred-stringy}.
Given this metric (\ref{D0metric}), 
the boundary ($r\to \infty$) is formally located at $z\sim 0$. 
One should keep in mind, however, that
the near-horizon condition $r^7\ll q$ in the original metric 
(\ref{orginal-D0metric}) puts a natural cutoff around $z\sim q^{1/7}$.

As in the standard AdS-CFT correspondence,
the time scale $\Delta t$, 
with which we probe the gauge theory,
corresponds to the radial 
scale $\Delta z\sim \Delta t$ from the bulk viewpoint, since 
 the time is common to both 
the bulk and boundary theories. 
(We will see this more explicitly in the following calculations.)
This relation enables us to convert the region of validity
(\ref{validity}) of the supergravity analysis
into 
\begin{equation}
\lambda^{-1/3} \ll \Delta t \ll \lambda^{-1/3}N^{10/21}
\label{timeregion}
\end{equation}
in terms of the time separation in the $(0+1)$-dimensional gauge theory,
where we have used
\beq
\lambda \equiv g_{{\rm YM}}^{\, 2} N = (2 \pi)^{-2} g_s N \ell_s ^{\, -3}  \ .
\eeq
This region (\ref{timeregion}) with milder $N$ dependence $N^{10/21}$ 
does not overlap on the IR side with the 
M-theory regime $\Delta t \sim \lambda^{-1/3} N$
mentioned below eq.~(\ref{Ham-Pminus}).
However, the Monte Carlo data presented 
in section \ref{sec:MCresults}
suggest that the supergravity results may be valid 
even in the IR region $\Delta t \gtrsim  \lambda^{-1/3}N^{10/21}$ 
beyond
the naive region of validity (\ref{timeregion})
at least for supergravity modes.


We emphasize that the whole near-horizon background 
including the Weyl factor, the dilaton and the RR 1-form $A_0dt$ 
(up to an irrelevant constant term for the 1-form) 
is invariant under the same generalized scaling symmetry 
that appears
on the gauge theory side, 
where we transform the parameter $g_s\to a^{3}g_s$ together with 
the coordinates $t\to a^{-1}t$, $z\to a^{-1}z$. We can therefore 
classify the linearized waves in supergravity around the near-horizon background 
according to the transformation properties under the GCS. 
Although the behavior of massive fields such as stringy excited modes 
is qualitatively different from the massless case, 
we can still obtain constraints from this scaling property.
See also ref.~\cite{Azeyanagi:2008mi}
for a discussion on the extension of this symmetry to 
the stringy level in a slightly different but related context.

\subsection{predictions for supergravity modes}
\label{sec:prediction-sugra}

Let us now review
the calculation of correlation functions 
from the supergravity side \cite{SY}. 
The basic idea is to apply the general prescription~\cite{GKPW} 
for the gauge-gravity correspondence,
which has been tested by many nontrivial examples 
in the standard AdS case. 
The prescription, formulated in the Euclideanized space-time, 
states that the correlation
functions of Matrix theory operators ${\cal O}^I(t)$ 
can be calculated by evaluating the bulk supergravity action
$S_{\rm SG}$ as a functional of the values of the 
bulk fields at the boundary as
\begin{equation}
e^{-S_{\rm SG}[h]} \Big|_{h^I=h^I_0} 
= 
\Big\langle e^{
\sum_I\int dt\,  h^I_0(t){\cal O}^I(t)} \Big\rangle \ .
\label{GKPW}
\end{equation}
The supergravity fields $h^I$ 
classified by an appropriate basis $\{ I\}$ are 
required to take a fixed value 
$h^I=h^I_0(t)$ at a {\it regulated} boundary $z=\epsilon$. 
In our case it is natural to assume that the boundary 
is located at the end of the near-horizon region
$\epsilon=q^{1/7}$ (or $r=q^{1/7}$).
The fields are assumed to vanish 
in the central region ($z\to \infty$).
This condition\footnote{From a mathematical point of view, 
it is needed for the kinetic operators 
of linear perturbations around the singular background
to be self-adjoint near the center.}
enables us to circumvent the singularity at the center,
and plays a crucial role in the following calculations.
The right-hand side of (\ref{GKPW})
represents the generating functional for 
connected correlation functions of Matrix theory
operators ${\cal O}^I(t)$, each of which couples to $h^I_0(t)$.  

In ref.~\cite{SY}, the complete 
spectrum of the linearized fluctuations around the
near-horizon D0-brane background was obtained. After diagonalizing
the fluctuations and decomposing them into SO(9) spherical harmonics, 
each eigen-mode of the bosonic fluctuations is described by the 
(Euclidean) effective action  
\begin{equation}
S_{{\rm eff}}={q\over \kappa^2}\int dt \, dz\, z 
\left( (\partial_0 h)^2+(\partial_z h)^2
+{\nu^2\over z^2}h^2
\right) \ .
\label{action}
\end{equation}
The field $h$ is normalized in such a way that they are
dimensionless and the action has an overall factor of $1/\kappa^2$,
where $\kappa^2\sim g_s^{\, 2} \, \ell_s^{\, 8}$ 
represents the gravitational coupling. 
The constant $\nu$ is given by 
\begin{equation}
\nu={2\over 5}\, \ell +{7\over 5} \, n \ ,
\label{nu-ell-n}
\end{equation}
where $\ell$ is an integer which represents the SO(9) total angular 
momentum, and $n$ is an integer which depends on the type of 
the field $h$ ($-1\le n\le 3$
for bosonic modes). 
In Table~\ref{dictionary} 
we list the modes which 
we study
by Monte Carlo simulation 
in section 3 (They are denoted in ref.~\cite{SY} 
as $v_2$, $a_2$ and $s_3$, respectively.),
and describe their properties such as
the transformation property on $S^8$, 
the value of $\nu$, and the range of $\ell$. 
The origin of these fields are 
either the graviton $g_{MN}$ or the three-form potential $C_{MNP}$
in 11D supergravity.
The corresponding operators in Matrix theory will be specified later
as (\ref{currents1})-(\ref{currents3}).
For a complete list of the spectrum, see ref.~\cite{SY}. 

\TABULAR[h]{|c|c|c|c|c|c|} {
\hline
mode & 11D origin & $S^8$ representation& $\nu$ & range of $\ell$
& operator\\
\hline
$v_2$&$C_{MNP}$& anti-sym. 2-form& 
${2\over 5}\ell$& $\ell\ge 1$&
$J_{\ell}^+$\\
\hline
$a_2$&$G_{MN}$& vector&${2\over 5}\ell$&$\ell\ge 2$& 
$T_{\ell}^+$\\
\hline
$s_3$&$G_{MN}$& scalar&${1\over 5}(2\ell-7)$&$\ell\ge 2$&
$T_{\ell}^{++}$\\
\hline
}
{Supergravity modes and the corresponding operators in Matrix theory.
\label{dictionary}}

Let us compute the two-point correlation functions corresponding 
to the supergravity modes
following the prescription in ref.~\cite{GKPW} 
(in particular, the first paper).
Imposing the boundary condition 
$h_0(t)=\int dp\, e^{ip t} \, \tilde{h}(p)$ 
at $z=\epsilon$,
one can write the solution to the wave equation 
obtained from the effective action (\ref{action}) as
\begin{equation}
h(t,z)=\int^{\infty}_{-\infty} dp \, e^{ip t} \, \tilde{h}(p) \, 
\frac{K_{\nu}(p z)}{K_{\nu}(p \epsilon)} \ ,
\label{solution}
\end{equation}
where $K_{\nu}(p z)$ represents the modified Bessel function 
of the second kind, which decays exponentially as $z\to \infty$. 
Upon substituting the solution (\ref{solution}), 
the action (\ref{action}) can be reduced to a boundary term.
Setting the boundary at $z=q^{1/7}$, and expanding the
action in powers of $p\, q^{1/7}$, we obtain 
\begin{eqnarray}
S&=&{q\over \kappa^2}\int dt \, 
\Bigl[zh\partial_z h
\Bigr]_{z=q^{1/7}}^{z=\infty} \nonumber \\
&=&{q\over \kappa^2}\int dp \, \tilde{h}(p) \, \tilde{h}(-p)
\left[
({\rm analytic\ in\ }p)+c \, \Bigl(p\, q^{1/7} \Bigr)^{2\nu}
\left\{ 1+ {\rm O}\Bigl((p\, q^{1/7})^2\Bigr)\right\} \right] \ , 
\label{action-eval-sol}
\end{eqnarray}
where $c$ is a non-zero constant factor.\footnote{This factor 
gives the normalization of the two-point function, 
but we leave it as a free parameter in the following analysis.
The correct normalization of some operators, 
such as the energy-momentum tensor, 
can be fixed by computing two- and three-point functions 
and requiring the consistency with the Ward identities. 
This was done in ref.~\cite{Kanitscheider:2008kd} using an expansion in 
the Fefferman-Graham form. We thank
K.~Skenderis for discussions on this point.}
The terms analytic in $p$ correspond to local divergences
(delta functions and its derivatives) in the coordinate space.
We ignore these terms
since they do not contribute to the correlation 
functions at finite separations, and furthermore
we keep only the leading terms which are 
most relevant to the IR property of the correlation 
functions.  
From the relation (\ref{GKPW}), we obtain 
the connected part of the two-point function at large separations as
\begin{eqnarray}
\label{OOpred}
\Bigl\langle {\cal O}(t)\, 
{\cal O}(t') \Bigr\rangle&=&\int dp \int dp' \, 
e^{ip t} \, e^{ip' t'}
{\delta\over \tilde{h}(p)}{\delta \over \tilde{h}(p')}S  \\
& \sim & 
{q^{1+{2\over 7}\nu}\over \kappa^2}\int dp \,
e^{ip(t-t')} \, p^{2\nu} 
\sim
{1\over \kappa^2} {q^{1+{2\over 7}\nu} \over |t-t'|^{2\nu+1}} \ ,
\label{sugraresult}
\end{eqnarray}
which leads to the result (\ref{coordspace}).

In the last step of (\ref{sugraresult}),
we have to recall that 
in order to have a well-defined Fourier transform
$\int_{-\infty}^{\infty} dp \, F(p) \, e^{ipx}$,
the $L^1$ integrability condition 
$\int_{-\infty}^{\infty} dp \, \Big|F(p)\Big|<\infty$
must be satisfied.
As usual, the nonintegrability for $|p|\rightarrow \infty$ 
can be treated as distributions, which 
have appreciable supports only at short distances with 
respect to time separation and hence can be ignored for 
studying the IR behaviors. 
On the other hand, since our correlation functions behave 
as $\sim |p|^{2\nu}$ at small $p$, the inverse Fourier 
integral for the operators with $\nu < -\frac{1}{2}$
(\emph{e.g.}, $T^{++}_2$ defined below) 
is divergent at $p\sim 0$.
In these cases, we would have to invoke analytic continuation 
with respect to the exponent $\nu$ 
in defining the Fourier integral. The resulting formula is well known 
($t\ne 0$), 
\beq
\frac{1}{\sqrt{2\pi}}\int_{-\infty}^{\infty}dp \, 
|p|^{2\nu}\, e^{ipt} =2^{2\nu+1/2}\, \frac{\Gamma(\nu+1/2)}{
\Gamma(-\nu)} \, |t|^{-2\nu-1} \ .
\label{ftransform}
\eeq
Since the predictions are not trustable in the UV region, 
we should use this formula only for sufficiently large $|t|$.

In the above calculation the information 
near the boundary plays a decisive role 
in discriminating various operators
since the boundary condition at the center
is chosen such that the fields should vanish.
All we need to know is the way in which the wave function is ``reflected'' 
from the central region, where
the wave function (\ref{solution}) decays 
exponentially as $z\to\infty$.
This vanishing behavior near the central strong coupling 
regime may be the reason why our result (\ref{sugraresult}) 
can be valid in the IR region beyond the naive region of 
validity (\ref{validity}). 
On the other hand, it should also be emphasized that 
while these calculations are performed totally within the 10D picture, 
the waves reflected back to the boundary 
can 
elicit important information 
by sinking into the near-central region during the reflection, 
where the dilaton grows and hence 
the 11th direction begins to open up indirectly. 
%
The further we go to the IR region of the gauge theory, the longer the corresponding 
waves stay near the central region. 
Consequently the reflected waves in the IR region at the boundary can in principle store 
richer traces of the 11th direction than in the UV region.

Let us now describe how to
find the operators in Matrix theory
which correspond to the supergravity modes discussed above.
For that purpose it turns out to be useful
to study the currents and their moments in Matrix theory 
within perturbation theory around appropriate backgrounds 
for the matrix variables following ref.~\cite{Taylor}.
These operators are written in a single trace form, and 
are coupled to gravitational perturbations.
One can identify them
by analyzing the one-loop effective potential 
between diagonal blocks, which represents gravitational interactions. 
The moments are constructed from the basic operators 
by inserting the $X_i$ fields
inside the trace and symmetrizing their ordering. 
Following the 11D light-cone notation in ref.~\cite{Taylor},
let us define the operators\footnote{%
The operators given in ref.\cite{Taylor} are of the form
$J_{ij;i_1\ldots i_\ell}^{+(\ell)}={1\over g_s\ell_s}
{\rm Tr}
\left(F_{ij}X_{i_1}\cdots X_{i_\ell}\right)$.
Here we have taken special combinations $J^{+}_{\ell}$,
which transform irreducibly under SO(9). The same remark applies to
the operators $T^{+}_{\ell}$ and $T^{++}_{\ell}$ as well.}
\begin{eqnarray}
J^{+}_{\ell}&=&{1\over g_s\ell_s} \, C_{ij;i_1\ldots i_\ell} \, {\rm Tr}
\left(F_{ij}\tilde{X}_{i_1}\cdots \tilde{X}_{i_\ell}\right) \ ,
\label{currents1}\\
T^{+}_{\ell}&=&{1\over g_s\ell_s} \, C_{i;i_1\ldots i_\ell} \, {\rm Tr}
\left((D_tX_{i})\tilde{X}_{i_1}\cdots \tilde{X}_{i_\ell}\right) \ ,
\label{currents2}\\
T^{++}_{\ell}&=&{1\over g_s\ell_s} \, C_{i_1\ldots i_\ell} \, {\rm Tr}
\left(\tilde{X}_{i_1}\cdots \tilde{X}_{i_\ell}\right) \ ,
\label{currents3}
\end{eqnarray}
where $F_{ij}=- i \, [X_i,X_j]/\ell_s^{\, 2}$,
which we will identify with the supergravity modes $v_2$, $a_2$, $s_3$
in Table~\ref{dictionary}, respectively.
We have defined the dimensionless matrix variable
$\tilde{X}_i=X_i/q^{1/7}$,
which naturally corresponds to the dimensionless combination 
$r^7/q$ appearing in the D0-background in the bulk theory.  
We have also assumed the same global prefactor $1/g_s\ell_s$ 
as in the D0-brane action (\ref{Matrixaction}) so that
the engineering dimensions of these operators are now 1.   
It turns out that this normalization is necessary
for matching the bulk modes and the gauge-theory operators 
with respect to the GCS. 
%

In order for the operators to have definite SO(9) angular momenta,
the constant coefficients $C$'s must satisfy the following conditions.
The coefficient $C_{i_1\ldots i_\ell}$ in (\ref{currents3})
should be totally symmetric, and it should also be traceless under
contraction of any two indices.
The coefficient $C_{i;i_1\ldots i_\ell}$ in (\ref{currents2}) 
should be totally symmetric, and it should also be
traceless with respect to the indices $i_1\ldots i_\ell$ ,
and anti-symmetric under the exchange of $i$ and 
any of $i_1\ldots i_\ell$.
The coefficient $C_{ij;i_1\ldots i_\ell}$ in (\ref{currents1}) 
is totally symmetric, and it should also be
traceless with respect to the indices $i_1\ldots i_\ell$,
and anti-symmetric in $i,j$ as well as for the exchange of 
$i$ or $j$ with any of $i_1\ldots i_\ell$. 
%
%

The guiding principle in relating the gauge-theory operators 
with the supergravity modes is that the
generalized conformal dimensions should match between the bulk
fields and the gauge-theory operators. 
Let us define
the generalized conformal dimension $\Delta$ of
a gauge-theory operator $O(t)$ such as (\ref{currents1})-(\ref{currents3})
by the scaling property
$O(t) \rightarrow O'(t')=\rho^{\Delta} \, O(t)$ 
under $t\rightarrow t'=\rho^{-1} \, t$, 
$g_s\rightarrow g_s'=\rho^3 \, g_s$. 
Having in mind the calculation from the gravity side
based on the effective action (\ref{action}),
we assume that the two-point functions of these operators obey 
the power-law behavior,
and 
that the only length 
scale allowed is $q^{1/7}$ apart from the 
gravitational constant $g_s^{\, 2} \, \ell_s^{\, 8}$ appearing 
as the overall coefficient. 
Then the GCS
along with the usual dimensional analysis
fixes the behavior of
the correlation function 
with the above normalization 
to have the general form
\begin{equation}
\Big\langle
{\cal O}(t) \, {\cal O}(0)
\Big\rangle
\sim \frac{1}{g_s^{\, 2} \, \ell_s^{\, 8}} \, q^{(\Delta + 6)/5}
\, |t|^{-(7\Delta +12)/5} \ .
\end{equation}
Comparing this with eq.~(\ref{sugraresult}), we get
$\Delta=-1+{10\over 7}\nu$.
We can see from Table~\ref{dictionary} 
and (\ref{currents1})-(\ref{currents3})
that 
$v_2$, $a_2$, $s_3$ 
correspond to $J^{+}_{\ell}$, $T^{+}_{\ell}$, $T^{++}_{\ell}$,
respectively, since the generalized conformal dimensions $\Delta$ 
coincide 
if we assign the canonical value 1 to $X_i$ as 
suggested from eq.~(\ref{GCS}) without any anomalous dimension.
(See ref.~\cite{SY} for a complete dictionary between the 
supergravity modes and the Matrix theory currents.) 
Note that each of $\tilde{X}_i=X_i/q^{1/7}$ contributes 
$1-{3\over 7} = {4\over 7}$
to the scaling dimension,
and hence ${2\over 5}$ to the index $\nu$ in accord with
(\ref{nu-ell-n}).
It is remarkable that the scaling properties can be explained 
with such simple assignment of dimensions 
with respect to the GCS.
We emphasize that from a purely gauge-theoretical point of view,
the appearance of the factor $q^{1/7}$ is 
genuinely a dynamical effect, which is difficult to understand 
without invoking the dual gravity theory.

As has been mentioned below eq.~(\ref{GCS}),
the generalized scaling transformation used above 
is essentially equivalent to the boost 
transformation along the 11th direction,
where the longitudinal momentum $P^+=N/R$ 
is scaled by treating $1/R$ as a variable with fixed $N$.
In the M-theory interpretation of the gauge theory, 
on the other hand, we fix $R$ or $g_s$ instead and 
increase $N$ to realize the IMF. 
It is therefore interesting to examine the above general form 
from this point of view. 
We consider the transformation
$N\rightarrow \rho \, N$ 
together with $t\rightarrow \rho \, t$.
Then the two-point functions (\ref{sugraresult}) scale as 
\begin{equation}
\Bigl \langle {\cal O}(t)\, {\cal O}(t') \Bigr \rangle
\sim  {1\over g_s^{\, 2} \, \ell_s^{\, 8}}
 {(g_s N \ell_s^{\, 7})^{1+{2\over 7}\nu} \over |t-t'|^{2\nu+1}}
\to  \rho^{-{12\nu\over 7}}
\Bigl \langle {\cal O}(t) \, {\cal O}(t') \Bigr\rangle \ ,
\label{two-pt-Mtheory}
\end{equation} 
from which we obtain 
the weight $d_{{\rm M}}=-6\nu/7$ for each operator 
under the M-theory boost.
In terms of the 11D light-like coordinates, 
the exponent $\nu$ is expressed as 
\begin{equation}
\nu={7\over 5} \, (1-n_++n_-)
    +{2\over 5}\, \ell \ ,
\label{nu-Mtheory}
\end{equation}
where $n_{+}$ $(n_{-})$ is the number of upper $+$
($-$) light-cone indices in the operator. 
Thus the weight $d_{{\rm M}}$
%
is given by
\begin{equation}
d_{\rm M}=
\left( 1+{1\over 5} \right)
(n_{+} - n_{-} -1)
- \left( {1\over 7}+{1\over 5} \right) \ell \ .
\end{equation}
This should be compared with the {\it kinematical} weight for the boost 
$d_{\rm M}^{\rm (kin)}= (n_{+} - n_{-} -1) - {1\over 7} \ell$,
where the term $-1$ in the parenthesis
comes from the fact that the currents
are supposed to be integrated over the $x^{-}$ direction,
and the factor of $1/7$ in front of $\ell$
comes from
our normalization of transverse fields  
$\tilde{X}_i=X_{i}/q^{1/7}$. 
The weight $d_{M}$ found from the gauge-gravity correspondence
is indeed determined solely from the 11-dimensional index structure
of the operator, but we observe interesting anomalous factors. 
It is therefore important to clarify whether 
the behavior (\ref{sugraresult})
continues to be valid in the 
M-theory regime corresponding to the far IR 
region $|t-t'|\propto N$. Our Monte Carlo data presented
in section \ref{sec:res-sugra}
seem to suggest that it does. 
Then, 
it would be interesting to clarify
the meaning of the anomalous behavior 
indicating that the transverse size is compressed 
by the factor of $\rho^{1/5}$ under the boost. 
%
See ref.~\cite{Y} for further considerations on this issue.

\subsection{predictions for stringy excited modes}
\label{sec:pred-stringy}

In this subsection we extend the calculation of
correlation functions to operators that correspond to
stringy excited modes on the gravity side.
We note first that this is a highly nontrivial issue
since one has to somehow generalize the GKPW prescription, 
which is based on the supergravity approximation.
In the most general case, one would have recourse to 
superstring field theory for closed strings, 
which, however, has never been formulated successfully in non-trivial 
classical backgrounds even at the linearized level. 
This forces us to formulate the gauge-gravity correspondence
in the first-quantized picture. 
Here we review a bulk analysis including stringy excited states \cite{ASY}. 
The analysis treats a single string moving in the 
S$^8$ direction with $J$ units of angular momentum 
with a small but finite excitation number. 
We can then use a semi-classical approximation 
for the center-of-mass motion of the string 
by taking the large-$J$ limit, 
which is often called the plane-wave limit. 
%
%

Following the well-known work
in the ${\rm AdS}_5\times {\rm S}^5$ case~\cite{BMN}, 
we identify the string states in the plane-wave limit with 
the gauge-theory operators 
which are the counterparts of 
the so-called BMN operators 
in the ${\cal N}=4$ SYM.
As a typical operator, let us consider
\begin{equation}
{\cal O}_{ij,n}^J 
=\sum_{k=0}^{J} e^{2\pi i kn/J} \, {\rm Tr} \left(
X_i Z^k X_j Z^{J-k} \right) \ ,
\label{typ-ope}
\end{equation}
with $Z=X_8+i X_9$, where the $X_8$ and $X_9$ are 
taken to be the directions of classical trajectories in S$^8$. 
The ``impurities''
$X_i$ and $X_j$ are the fields in the other transverse directions
($i,j=1,\cdots , 7$). 
The operator (\ref{typ-ope})
corresponds to a state
with two oscillators at the $n$-th level being excited.
Supergravity modes discussed in the previous subsection
correspond to the operators with $n=0$.
We can also consider excitation of 
oscillators in the $t$ direction, which corresponds to an operator
with the covariant derivative $D_{t}$ inserted as an impurity.

The basis of this analysis is the prescription \cite{DSY}
introduced originally to solve some puzzles
in the original BMN proposal. 
More recently, this method\footnote{For a compact review of this approach, 
see ref.~\cite{Yoneya:2006td}.} has been 
applied systematically to the evaluation of various 
correlation functions including three-point functions 
in the ${\rm AdS}_5\times {\rm S}^5$ case \cite{Dobashi:2004nm, Tsuji}. 
Instead of computing the energy of strings in the global AdS and 
identifying them with the scaling dimensions 
as in ref.~\cite{BMN}, 
one computes the transition amplitude along a trajectory 
in the Euclidean AdS 
(with a non-trivial Weyl factor in the D0 case) 
connecting two points at the boundary. 
Note that the real trajectories 
in the Minkowskian AdS, in general, 
do {\it not} reach the boundary, while the tunneling 
trajectories in the Euclidean AdS start and end 
at the boundary. 
The amplitudes along the tunneling trajectories 
can then be identified with the gauge-theory 
correlation functions.  
This 
gives us a definite way to incorporate stringy excitations.
In the present D0 case, in particular,
it is crucial to use this approach since 
the original BMN approach would require 
us to consider the strings moving in a singular region near the center. 
That the tunneling trajectories do not reach the singularity 
at the center corresponds to 
the boundary condition in the supergravity analysis
that the wave function should vanish at the center.

We now review the key steps of this approach. 
For simplicity, we set $\alpha'(=\ell_s^{\, 2})=1$ 
throughout this subsection. Although the original analysis was 
given for general D$p$-branes with $p<5$, we restrict ourselves
to the $p=0$ case, which is relevant to the present work.
To explain the idea in a simple setting, let us first start by considering
a geodesic for an ordinary massive particle on the Euclidean ${\rm AdS}_2$,
which can be obtained by minimizing the action
\begin{equation}
S=m\int d\tau\sqrt{\tilde{g}_{\mu\nu}\partial_\tau x^{\mu}
\partial_\tau x^{\nu}} \ ,
\label{particleaction}
\end{equation} 
where $\tilde{g}_{\mu\nu}$ is the metric for Euclidean ${\rm AdS}_2$
given by $ds^2=(dt^2+dz^2)/ z^2$. 
If we set the mass\footnote{The factor of $2/5$ in the mass $m$ 
is due to the fact that in the D0-background, the
radius of S$^8$ is $5/2$ in units of the AdS radius.
Note also that we have ignored the $J$-independent term in the mass $m$
since we are dealing with the large-$J$ limit.   
} to $m={2\over 5}J$, we can regard this action (\ref{particleaction})
as that of a Kaluza-Klein particle with $J$ units of
angular momentum on the compactified S$^8$
with the time coordinate being Euclideanized.
The geodesic of our interest 
is a half circle\footnote{Here we take the 
coordinates $(t,z)$ to be dimensionless.
In order to retrieve the original coordinates, 
which scale like (\ref{GCS}) under generalized conformal
transformation, one has to multiply them by $q^{1/7}$.}
\begin{equation}
t=\tilde{\ell} \, \tanh \tau \ ,\quad 
z={\tilde{\ell}\over \cosh\tau} \ ,
\label{geodesic}
\end{equation}
which connects two points on the boundary
separated by $|t_f -t_i|= 2 \, \tilde{\ell}$. 
One finds from (\ref{geodesic}) that
the radial cutoff $(z>\epsilon)$ is
related to the proper time cutoff $(|\tau|<T)$ as 
\begin{equation}
\epsilon\sim 2 \, \tilde{\ell} \, e^{-T}
= |t_f-t_i| \, e^{-T} \ .
\end{equation}
Substituting this solution into the action (\ref{particleaction}),
we get $S_{\rm cl}=m\int_{-T}^{T} d\tau=2mT$. The amplitude
in this geodesic approximation is then proportional to 
\begin{equation}
e^{-S_{\rm cl}}=e^{-2mT}=\left({\epsilon \over |t_f-t_i|}\right)^{{4\over 5}J}.
\label{classicalamplitude}
\end{equation}
This gives the large-$J$ limit of the two-point function 
for the operator ${\rm Tr} \Bigl(Z^{J} \Bigr)$,
which is included in (\ref{currents3}) as a particular case.
The $J$ dependence is indeed consistent\footnote{The $J$-independent 
term in the exponent can be obtained
by computing the zero-point energy of the superstring
as described in the second paper of ref.~\cite{ASY}, 
and it also agrees with the result from the GKPW prescription.}
with the result (\ref{sugraresult}) obtained using the wave picture 
for the operator $T_{\ell}^{++}$ with $\ell = J$.

In order to treat more general operators, we need to embed 
the above point-like classical trajectory in string theory.
For that purpose we first perform a {\it double} Wick rotation~\cite{DSY} 
$t\to -it$, $\psi\to -i\psi$
in the D0-brane metric (\ref{D0metric}). One can check that the classical trajectory discussed above 
for the Euclideanized ${\rm AdS}_2$ case 
agrees\footnote{The difference of the Weyl factor 
in the background metric can be
absorbed, along the classical trajectory, 
by a time-dependent reparametrization of $\tau$.}, 
in the two-dimensional 
subspace of time and radial coordinates, with a point-like ground-state solution for 
the standard (so-called ``Polyakov-type'') worldsheet action 
for strings around 
the D0-background satisfying the Virasoro conditions.
The solution is independent of the 
worldsheet coordinate $\sigma$ and 
is given by eq.~(\ref{geodesic}) and $\psi={2\over 5} \tau$.
The period of $\sigma$ is chosen to be $0\le \sigma\le 2\pi \alpha$ 
with $\alpha=5J/(2q^{2/7})$ so that the canonical momentum for $\psi$
is equal to $J$, \emph{i.e.}, 
$J={1\over 2\pi}q^{2/7}\int d\sigma \dot{\psi}$. 
The semi-classical amplitude (\ref{classicalamplitude}) can then be
reproduced by evaluating the Routh function for the worldsheet 
string action along this solution. 
%

The next task is to expand the worldsheet action 
around this classical solution to the quadratic order in 
fluctuations. The quadratic terms are given as\footnote{We have 
fixed the worldsheet metric as 
$\sqrt{h}h^{\tau\tau}=(\sqrt{h}h^{\sigma\sigma})^{-1}=r^{3/2}(\tau)$,
as in the second paper of ref.~\cite{ASY}.
With this choice, $m_x$, $m_y$ are constant, but $x'^2$, $y_i'^2$
get $\tau$ dependent coefficients. 
If we use the conformal gauge as in the first paper of ref.~\cite{ASY}, 
$m_x$, $m_y$ become $\tau$-dependent. 
Final results do not depend on the gauge choice.}
\begin{equation}
S^{(2)}={1\over 4\pi}\int d\tau\int_{0}^{2\pi\alpha} d\sigma \, 
\Bigl( \dot{x}^2+r^{-3}(\tau)\, x'{}^2+m^2_x \, x^2
+\dot{y}_i^2+r^{-3}(\tau) \, y_i'{}^2+m^2_y \, y_i^2 \Bigr) \ , 
\label{worldsheet-action-quad}
\end{equation}
where $r(\tau)$ represents the trajectory in the radial direction
\begin{equation}
r(\tau)=\left( \frac{2\cosh\tau}{ 5 \, \tilde{\ell}} \right)^{2/5} \ ,
\end{equation}
which is obtained from (\ref{geodesic}) using (\ref{poincare-coord}).
The fields $x$ and $y_i$ $(i=1,\cdots, 7)$
in (\ref{worldsheet-action-quad})
represent fluctuations within the ($t,z$) direction
and along the S$^8$ direction, respectively.
These fields have mass $m_x=1$ and $m_y=2/5$ due to the
curvature around the geodesic. 

The supergravity modes can be obtained by restricting ourselves
to point-like configurations, which corresponds to setting
$x ' = y_i ' =0$ in the worldsheet action (\ref{worldsheet-action-quad}).
Then the problem reduces to that of ordinary harmonic oscillators 
with the Hamiltonian 
$H=m \Bigl(a^\dagger a+ \frac{1}{2} \Bigr)$
written schematically for each of the $x$ and $y_i$ oscillators.
The amplitude from $\tau=-T$ to $\tau=T$ is given by
\begin{equation}
e^{-2HT}
=e^{-2m(a^\dagger a+1/2)T}
=\left(\epsilon\over |t_f-t_i| \right)^{2 m(a^\dagger a+1/2)} \ .
\label{ordinary-HOC}
\end{equation}
An insertion of $X_i$ ($i=1,\cdots, 7$) into the ground state
operator 
${\rm Tr} \Bigl(Z^{J} \Bigr)$
corresponds to exciting
one of the $y_i$ oscillators with $m=m_y$, which 
increases the power of $|t_f-t_i|$ 
in the two-point function by $2 m_y = 4/5$.
Similarly, an insertion of $D_t$ corresponds
to exciting the $x$ oscillator with $m=m_x$,
which increases the power by $2 m_x = 2$. 
These rules are consistent with the results 
from the wave-function analysis shown in Table~\ref{dictionary}.  
For instance, let us consider the operator $T_{\ell}^{++}$ 
with $\ell=J$ and with all of the transverse directions
confined in the $Z$-plane. Inserting $D_t$ to it 
creates an operator $T_{\ell}^{+\, i}$ with $\ell=J-1$.
The value of $\nu$ increases by 1 as one can see from Table~\ref{dictionary},
which implies that the power increases by 2 in accord with the
conclusion obtained from (\ref{ordinary-HOC}).
%
Note that, in the present tunneling picture, the 
time direction $t$ of gauge theory  
is transverse to the $\tau$-direction 
near the boundary.  As a consequence, we can naturally 
identify $D_t$ with oscillation in a transverse direction,
which supports the consistency of the present analysis.
In the Minkowskian picture, this correspondence is unclear.

In order to study stringy excited modes, 
we need to allow $x '$ and $y_i '$ to be non-zero
in eq.~(\ref{worldsheet-action-quad}).
Since the $x'{}^2$ term and the $y_i'{}^2$ term have 
$\tau$-dependent coefficients,
we have to solve the problem of 
quantizing oscillators with time-dependent masses.
Let us explain how this can be done
by taking one of the $y_i$ oscillators as an 
example.\footnote{Impatient readers may skip these details and 
go directly to the final result (\ref{stringamplitude}), 
which is valid for states with high wave numbers. }
 
The equation of motion for the $n$-th Fourier mode in $\sigma$ is
given by 
\begin{equation}
{d^2\over d\tau^2}\, y(\tau) = m^2(\tau)\, y(\tau) \ , \quad 
m^2(\tau)=r^{-3}(\tau)\, {n^2\over \alpha^2}+{4\over 25} \ .
\label{timedependenteom}
\end{equation}
Let the functions $f_{\pm}(\tau)$ be
the solutions to eq.~(\ref{timedependenteom}) 
with the boundary condition $f_{\pm}(\tau)\to 0$ 
as $\tau\to \pm\infty$, respectively, 
which are normalized by
$f_{+}\dot{f}_{-}-f_{-}\dot{f}_{+}=1$. 
Let us regard $y(\tau)$ now as an operator
and denote its conjugate canonical momentum by
$p(\tau)=i \, \partial_\tau y(\tau)$
satisfying the canonical commutation relation $[y,p]=i$.
Then we can write the solution to the time-evolution equation
(\ref{timedependenteom}) as
\begin{equation}
y(\tau)= f_{+}(\tau)\, a + f_{-}(\tau) \, a^{\dagger} \ ,
\end{equation}
where $a$ and $a^{\dagger}$ are
$\tau$-independent operators satisfying $[a,a^{\dagger}]=1$.
Due to the time reflection symmetry of the problem, 
we have 
$y^{\dagger}(\tau)=y(-\tau)$, 
which corresponds to 
the reality condition in the real-time formulation, 
and hence $f_{\pm}(\tau)=f_{\mp}(-\tau)$ 
since $f_{\pm}(\tau)$ are real in our case.

Using the Hamiltonian
$H(\tau)=\frac{1}{2} \Bigl( p^2+m^2(\tau)y^2 \Bigr)$,
we can define the transition amplitude 
between an initial state at $\tau=-T$ and a final state 
at $\tau=+T$ along the tunneling trajectory.
The corresponding ``Euclidean'' S-matrix is given by
$ S(T)={\cal T}\exp\left[-\int_{-T}^{T}d\tau H(\tau)\right] $.
Since the Hamiltonian is quadratic in $a$ and $a^{\dagger}$, 
the S-matrix takes the form
\begin{equation}
S(T)=N(T): \exp
\left({1\over 2}\, A(T) \,  (a^\dagger)^2 + B(T) \, a^\dagger a
+ {1\over 2} \, A(T) \, a^2 \right): \ ,
\end{equation}
where the coefficients are expressed in terms of $f_{\pm}(T)$ as 
\begin{equation}
A=-{1\over 2}\left({f_{+}(T)\over f_{-}(T)}
+{\dot{f}_{+}(T)\over \dot{f}_{-}(T)}\right) \ ,  \quad 
N^2=1+B={1\over 2f_{-}(T)\dot{f}_{-}(T)} \ .
\end{equation}
We can rewrite $S(T)$ as
\begin{equation}
S(T)=
{\cal N}(T)\exp\left(-\Omega \, b^\dagger b\right) \ ,
\end{equation}
where $b, b^\dagger$ satisfying $[b, b^\dagger]=1$ are
related to $a, a^\dagger$ through a $T$-dependent 
Bogoliubov transformation,
and $\Omega$ is given by
\begin{equation}
\cosh \Omega={1\over 2}\left(1+B+{1-A^2\over 1+B}\right) \ .
\end{equation}
In fact, with our boundary conditions for $f_{\pm}(\tau)$, 
it turns out that $A\sim (1+B)\sim 0$, $b \sim a$ 
and ${\cal N} \sim  (1 + B)^{1/2}$ when $T\to \infty$.
Thus the final result takes the simple form 
\begin{equation}
S(T) \sim (1+B)^{a^\dag a + 1/2} = 
\Bigl( 2 \, f_{-}(T) \, \dot{f}_{-}(T) \Bigr)^{- (a^\dag a + 1/2)} \ .
\label{final-result}
\end{equation}

Let us then discuss the behavior of stringy excited modes. 
Near the boundary ($r\to \infty$), the constant term $\frac{4}{25}$
in eq.~(\ref{timedependenteom}) is dominant over 
the time-dependent string mass term ($r^{-3}(\tau )n^2/\alpha^2$).
This results from the fact that
the curvature of the background in the string unit 
becomes strong as $r\to\infty$. 
However, when we study the large distance behavior 
$|t_f-t_i|\to \infty$, the geodesic passes deeply 
through the interior region, where the string mass term dominates. 
In this situation it is difficult to determine 
the functions $f_{\pm}(\tau)$ for general excited states. 
We therefore restrict ourselves to excited
states with higher wave numbers 
satisfying $n/\alpha\sim n \, q^{2/7}/J\gg 1$.
This allows us to neglect the terms other than 
the string mass term, 
considering that the regulated boundary is placed 
at the end of the near horizon region ($r(\tau)\sim 1$).
Therefore the equation of motion can be approximated as 
\begin{equation}
{d^2\over d\tau'^2} \, \tilde y={n^2\over \alpha^2} \, \tilde y \ ,
\end{equation}
where $\tau'$ is defined by $d\tau'/d\tau=r^{-3/2}$,
and $y=r^{3/4}\tilde{y}$. 
The solutions to this equation are 
\begin{equation}
f_{+}(\tau')=\sqrt{\alpha\over 2|n|} \, e^{-{|n|\over \alpha}\tau'} \ ,
\quad
f_{-}(\tau')=\sqrt{\alpha\over 2|n|} \, e^{{|n|\over \alpha}\tau'} \ .
\end{equation}
In terms of $\tau'$, the boundary $(r\to\infty)$ is reached at 
finite time $T_{\rm b}$, which is evaluated as
\begin{equation}
T_{\rm b} \sim \tilde{\ell}\int_{r_{\rm min}} ^{\infty} 
\frac{dr}{\sqrt{{25\over 4}\, \tilde{\ell}^{ \, 2} r^{5}-1} }
\sim \tilde{\ell}{}^{\, 3/5} \ . 
\end{equation}
Plugging this into eq.~(\ref{final-result}) with $T=T_{\rm b}$,
we obtain the S-matrix as
\begin{eqnarray}
S(T)
&\sim& \exp\left\{ -{2|n|\over \alpha} \, 
T_{\rm b} \left( a^\dagger a+ \frac{1}{2} \right) \right\} \nonumber \\
&\sim& \exp\left\{ -\hat{c}(n,  J) 
\left(a^\dagger a+\frac{1}{2}\right) \, 
q^{1/5}|t_f-t_i|^{3/5} \right\} \ , 
\label{stringamplitude}
\end{eqnarray}
where $\hat{c}(n, J)$ is proportional to $|n|/J$.
In the last equality, we have substituted $\tilde{\ell}$ in
$T_{\rm b}$ by $q^{-1/7}|t_f-t_i|$,
where $t_i$ and $t_f$ represent
the time coordinates on the boundary at both ends of the trajectory. 
Note that the combination 
$q^{1/5}|t_f-t_i|^{3/5}$ in the exponent is invariant under the 
generalized conformal transformation. 
In the case of general D$p$-branes ($p<5$), 
this combination takes the form $q^{1/(5-p)}|t_f-t_i|^{(3-p)/(5-p)}$. 

Note that, in the conformal case $p=3$, 
the power of $|t_f-t_i|$ in the exponent vanishes.
Therefore, the power-law behavior of correlation functions
is modified for stringy operators \cite{Dobashi:2004nm} 
only by the emergence of {\it anomalous} conformal dimensions,
which depend both on the Yang-Mills coupling constant and 
on the wave number $n$. 
In contrast to this, for the present non-conformal case, 
we found the peculiar exponential behavior (\ref{stringamplitude}) 
in the IR region. This is an interesting prediction,
which is worth being tested.
We consider that the simple form $|n|/J$ 
of the coefficient in front of the 
GCS invariant combination $q^{1/5}|t_f-t_i|^{3/5}$ 
on the exponent may be valid only 
in the present limit $n/\alpha \gg 1$. 
In general, it could happen that 
the GCS invariant combination 
$\Bigl( q^{1 /5}|t_f-t_i|^{3/5} \Bigr)^{\eta}$
appears on the exponent with some power $\eta \ne 1$.
In section \ref{sec:res-stringy} we will provide some evidence 
that the form (\ref{stringamplitude}) is actually valid 
for stringy operators even with small $J$ and $n$. 
Since the predicted behavior
is quite different from the standard behavior of 
a massive theory, 
we discuss the meaning of this form
based on its spectral representation in appendix \ref{sec:spectral-rep}.


\section{Monte Carlo calculations on the gauge theory side}
\label{sec:MCresults}

In the previous section we have reviewed the calculation
of two-point correlation functions
in Matrix theory from the gravity side
based on the gauge-gravity correspondence.
In particular, the power-law behavior (\ref{sugraresult}) 
was predicted
for the operators (\ref{currents1})-(\ref{currents3}),
which correspond to the supergravity modes
as summarized in~Table \ref{dictionary}.
We also have a prediction (\ref{stringamplitude})
for the type of operators (\ref{typ-ope})
corresponding to stringy excited modes 
in the large-$J$ and large-$n$ limits.
In this section we calculate these two-point correlation functions
directly on the gauge theory side by a Monte Carlo method.
We will see that our results for $N=2$ and $N=3$ already
show striking agreement with the predictions from the bulk side.

\subsection{putting Matrix theory on a computer}

We consider the Euclidean version of 
the action (\ref{Matrixaction}), which is given by 
\begin{eqnarray}
S = 
\frac{1}{g_{\rm YM}^{\, 2}} 
\int
d t \, 
{\rm Tr}  \, 
\left(
\frac{1}{2} \, (D_t X_i)^2 - 
\frac{1}{4} \, [X_i , X_j]^2  
+ \frac{1}{2} \, \psi_\alpha D_t \psi_\alpha
- \frac{1}{2} \, \psi_\alpha \gamma_i^{\alpha\beta} 
 [X_i , \psi_\beta ]
\right) \ .
\label{cQM}
\end{eqnarray}
Since the coupling constant $g_{\rm YM}^{\, 2}$ can be absorbed
by appropriate rescaling of the fields and the coordinate $t$,
we set the 't Hooft coupling $\lambda = g_{\rm YM}^{\, 2} N $ 
to unity without loss of generality.
Then the strong coupling limit amounts to the IR limit.\footnote{From 
the bulk point of view, once the near-horizon limit is assumed, 
the only spatial length scale is 
$\Delta X \sim (g_sN)^{1/3} \ell_s$, which
corresponds on the boundary to the temporal scale 
$\Delta t \sim (g_s N)^{-1/3} \ell_s \sim (g_{{\rm YM}}^{\, 2} N)^{-1/3}$.
These estimates of $\Delta X$ and $\Delta t$ 
are consistent with
the space-time uncertainty relation 
$\Delta t \, \Delta X \gtrsim \ell_s^{\, 2}$~\cite{Ystu}. 
}
Since the ${\rm U}(1)$ sector of the ${\rm U}(N)$ theory
is decoupled from the rest,
we actually study the ${\rm SU}(N)$ theory.

In order to put the theory on a computer,
we need to make the field degrees of freedom finite
by introducing UV and IR cutoffs in the $t$-direction.
The IR cutoff is introduced
by compactifying the $t$-direction to a circle of circumference
$\beta$. Since we are not interested in the thermal properties
of the system in this work,
we impose periodic boundary conditions 
on both bosons and fermions 
so that the supersymmetry is not broken by finite $\beta$ effects.
The UV cutoff is introduced
by a sharp cutoff in the momentum space
following the previous works \cite{Hanada-Nishimura-Takeuchi,%
AHanada-Nishimura-Takeuchi,Hanada-Hyakutake-Nishimura-Takeuchi,%
HMNT,HNSY}.
This has the following advantages over the conventional 
lattice approach \cite{Catterall:2007fp}.
Firstly the breaking of supersymmetry is milder
than in the lattice regularization, 
and it restores quite fast as the UV cutoff is 
removed \cite{Hanada-Nishimura-Takeuchi}.
Secondly one can implement the Fourier acceleration \cite{Catterall:2001jg}
without extra cost, which makes the simulation much more efficient.


When we introduce the momentum cutoff,
we have to take care of the gauge symmetry.
In the case of one-dimensional theory in general, 
one can fix the gauge completely
(\emph{i.e.}, choose a unique representative for each
gauge orbit non-perturbatively)
in such a way that higher momentum modes 
are more suppressed by the kinetic term in the action.
With this gauge choice, one can also show by simple power counting
that there is no UV divergence (due also to one dimension).
Therefore one can retrieve gauge invariant results 
by sending the UV cutoff to $\infty$
as is checked explicitly
for the bosonic theory \cite{Hanada-Nishimura-Takeuchi}.

Let us describe how we introduce
the UV cutoff more in detail. 
First we take the static diagonal gauge
$A(t) = \frac{1}{\beta} \, {\rm diag}
(\alpha_1 , \cdots , \alpha_N)$, 
where $\alpha_a$ can be chosen to satisfy the constraint 
$\max_a (\alpha_a) - \min_a (\alpha_a) 
< 2\pi$ 
by using the large gauge transformation with a non-zero winding number.
We have to add to the action a Faddeev-Popov term
\begin{eqnarray}
S_{\rm FP} =
- \sum_{a<b} 2 \ln 
\left| \sin \frac{\alpha_a - \alpha_b}{2}
\right|  \ , 
\end{eqnarray}
and the integration measure for $\alpha_a$
is taken to be uniform.
Having fixed the gauge completely by the above procedure, 
we can introduce a cutoff $\Lambda$
in the Fourier-mode expansion 
\begin{eqnarray}
X_i ^{ab} (t) = \sum_{n=-\Lambda}^{\Lambda} 
\tilde{X}_{i n}^{ab} \, e^{i \omega n t} \ , \qquad
\psi_\alpha ^{ab} (t) = \sum_{n=-\Lambda}^{\Lambda}
\tilde{\psi}_{\alpha n}^{ab} \, e^{i \omega n t}  \ ,
\label{Fourier-exp}
\end{eqnarray}  
where $\omega = 2 \pi / \beta$, taking into account that
we impose periodic boundary conditions on both $X_i ^{ab} (t)$ and 
$\psi_\alpha ^{ab} (t)$.
Using a shorthand notation
\begin{eqnarray}
\Bigl(f^{(1)}  \cdots  f^{(p)}\Bigr)_n 
\equiv \sum_{k_1 + \cdots + k_{p}=n}
f^{(1)}_{k_1} \cdots f^{(p)}_{k_p} \ ,
\end{eqnarray}
we can write the action 
(\ref{cQM}) as
$S=S_{\rm b}+S_{\rm f}$, where
\begin{eqnarray}
S_{\rm b}
&=&  N \beta
\Bigg[
\frac{1}{2} \sum_{n=-\Lambda}^{\Lambda} 
\left( n \omega - \frac{\alpha_a - \alpha_b}{\beta} 
\right)
^2   \tilde{X}_{i , -n}^{ba} \tilde{X}_{i n}^{ab}
- \frac{1}{4} \, 
{\rm Tr} \Bigl( [ \tilde{X}_{i} , \tilde{X}_{j}]^2  \Bigr)_0
\Bigg] \ , 
\label{bosonic_action} \\
S_{\rm f}
&=& \frac{1}{2} \, 
N \beta \Biggl[ \sum_{n=-\Lambda}^{\Lambda} 
i 
\left(
n \omega - \frac{\alpha_a - \alpha_b}{\beta}  
\right)
\tilde{\psi}_{\alpha , -n}^{ba} \tilde{\psi}_{\alpha n}^{ab} 
- (\gamma_i)_{\alpha\beta} \, 
{\rm Tr} \Bigl( \tilde{\psi}_{\alpha , -n} 
[ \tilde{X}_{i},\tilde{\psi}_{\beta}]  \Bigr)_0 \Biggr] \ .
\label{bfss_action_cutoff}
\end{eqnarray} 

Integrating out the fermionic variables, one obtains
the Pfaffian ${\rm Pf}{\cal M}$, which is complex
in general. (See eq.~(\ref{def-calM}) for the definition of ${\cal M}$.)
The phase of the Pfaffian turns out to be quite small for SU$(2)$
in the parameter region investigated in the present work,
but it does fluctuate for SU$(3)$, in particular at large $\beta$.
%
Here we simply neglect the phase and use $|{\rm Pf}{\cal M}|
= {\rm det} ( {\cal D}^{1/4})$, 
where ${\cal D}={\cal M}^\dag {\cal M}$, 
instead of ${\rm Pf}{\cal M}$. 
The system with finite degrees of freedom we arrive at in this way
can be simulated by using the
Rational Hybrid Monte Carlo (RHMC) 
algorithm \cite{Clark:2003na}, 
which has become quite standard nowadays 
in lattice QCD simulations with dynamical quarks.
The details of the algorithm are given in appendix \ref{sec:detail-MC}.


An important assumption in our method is
that the phase of the Pfaffian can be neglected.
While the previous results \cite{AHanada-Nishimura-Takeuchi,%
Hanada-Hyakutake-Nishimura-Takeuchi,HMNT,HNSY},
which confirmed the gauge-gravity correspondence with high accuracy,
certainly support this assumption, we cannot provide purely theoretical 
justification at the present stage. 
The effect of the phase can be incorporated in principle
by reweighting when one calculates expectation values.
This direct method becomes impractical due to huge cancellations
when the phase fluctuates violently.
We suspect that the fluctuation of the phase is actually 
smaller than that of typical extensive quantities
at large $\beta$ and that the phase-quenching is completely
justifiable in the large-$\beta$ limit.
In appendix \ref{sec:sign-problem}
we investigate the effect of the phase on a typical observable 
and show that it is indeed negligible.


\subsection{results for supergravity modes}
\label{sec:res-sugra}

In this subsection we present our results
for the two-point correlation functions of 
operators (\ref{currents1})-(\ref{currents3})
corresponding to supergravity modes.

Let us first define the operators we study by simulating
the model (\ref{cQM}).
The operators $J^{+ij}_{\ell,i_1,\cdots,i_\ell}$ ($\ell\ge 1$)
are defined by 
\begin{eqnarray}
J^{+ij}_{\ell,i_1,\cdots,i_\ell}
\equiv
\frac{1}{N} \, {\rm Str}
\Bigl(
F_{ij}X_{i_1}\cdots X_{i_\ell}
\Bigr) \ , 
\label{Jplus}
\end{eqnarray}
where $F_{ij}\equiv -i \, [X_i,X_j]$ and 
${\rm Str}$ represents the symmetrized trace treating $F_{ij}$ 
as a single unit. 
Restricting ourselves to $\ell \le 7$, we may assume 
that all the indices $i$, $j$, $i_1$, $\cdots$, $i_\ell$ are
different from each other so that the traceless condition 
for (\ref{currents1}) is trivially satisfied. 
The operators
$T^{++}_{\ell,i_1,\cdots,i_\ell}$ ($\ell\ge 2$) are defined by
\beq
T^{++}_{\ell,i_1,\cdots,i_\ell}
\equiv\frac{1}{N}\,  {\rm Str} 
\Bigl( X_{i_1}\cdots X_{i_\ell} \Bigr)  \ .
\eeq
Restricting ourselves to $\ell \le 9$, we may assume that
all the indices $i$, $j$, $i_1$, $\cdots$, $i_\ell$ are
different from each other so that the 
traceless condition 
for (\ref{currents3}) is trivially satisfied.
The operator $T^{+i}_{\ell,i_1,\cdots,i_\ell}$ are defined by 
\begin{eqnarray}
T^{+i}_{\ell,i_1,\cdots,i_\ell}\equiv
\frac{1}{N} \, {\rm Str}
\Bigl(
(D_t X_i)
X_{i_1}\cdots X_{i_\ell} \Bigr) \ , 
\label{Tplus}
\end{eqnarray}
where ${\rm Str}$ represents the symmetrized trace treating $(D_t X_i)$ 
as a single unit. 
Restricting ourselves to $\ell \le 8$, we may assume that
all the indices $i$, $i_1$, $\cdots$, $i_\ell$ are
different from each other so that the 
traceless condition 
for (\ref{currents2}) is trivially satisfied.
Due to the SO(9) symmetry, the result for the
two-point correlation function for each type of operators
should not depend on the assignment of the indices.
Therefore we average over all possible assignment 
to increase the statistics in actual calculation.

Since the basic dynamical degrees of freedom 
in our Monte Carlo calculations are Fourier modes 
(\ref{Fourier-exp}), the correlation functions that are
directly accessible are those in the momentum space
$\left\langle
\tilde{{\cal O}}(p) \, 
\tilde{{\cal O}}(-p)
\right\rangle$,
which are related to those in the real space
by the inverse Fourier transformation
\beq
\Bigl\langle
{\cal O}(t) \, {\cal O}(0)
\Bigr\rangle
=
\int \frac{dp}{2\pi}
\left\langle
\tilde{{\cal O}}(p) \,
\tilde{{\cal O}}(-p)
\right\rangle
e^{ipt} \ .
\label{inverse-Fourier-corr}
\eeq
Using the Fourier modes $\tilde{\cal O}_n$ 
defined similarly to (\ref{Fourier-exp}), we can rewrite it as
\beq
\left\langle
\tilde{{\cal O}}(p) \, 
\tilde{{\cal O}}(-p)
\right\rangle
= \beta \left\langle
\tilde{{\cal O}}_n \, 
\tilde{{\cal O}}_{-n}  
\right\rangle \ , \quad \mbox{where~}
p=\frac{2\pi n}{\beta} \ .
\eeq
Note that the factor of $\beta$ on the right-hand side
is needed to make the correlation function finite 
in the large-$\beta$ limit.

The gauge-gravity correspondence 
predicts 
the two-point function in the Fourier space to behave as 
\begin{eqnarray}
\Bigl\langle \tilde{{\cal O}}(p) \, \tilde{{\cal O}}(-p) \Bigr\rangle
\sim f(p) + g(p)\, |p|^{2\nu}  
\label{correlator_momentum_space} 
\end{eqnarray}
at small $p$ as one can see from 
eqs.~(\ref{action-eval-sol}) and (\ref{OOpred}).
Here, $f(p)$ and $g(p)$ ($g(0)\ne 0$) are analytic functions 
invariant under $p\leftrightarrow -p$, 
and hence they can be written as
$f(p)=f_0 + f_2 \, p^2 + \cdots$ and $g(p) = g_0 + g_2 \, p^2 + \cdots$,
where $g_0 \neq 0$.
The coefficients are not fixed 
since we have focused on
the most relevant term in the IR limit
neglecting the overall normalization factor.
Our main task is to extract the power $\nu$
by fitting the Monte Carlo data
for various correlation functions to the form
(\ref{correlator_momentum_space}) 
and to compare it with the values predicted by the
gauge-gravity correspondence.

    \FIGURE[b]{
    \epsfig{file=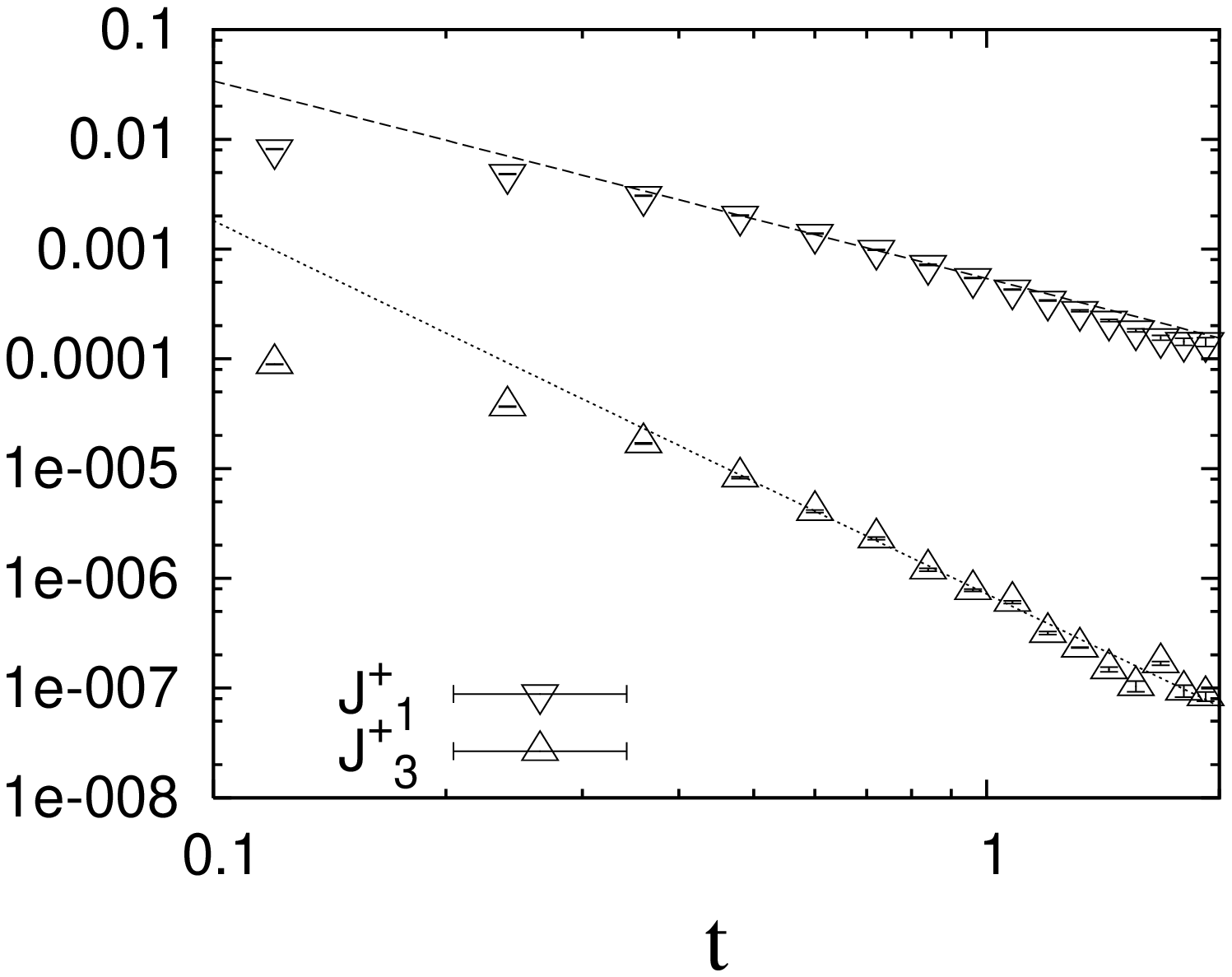,%
angle=270,width=7.4cm}
    \epsfig{file=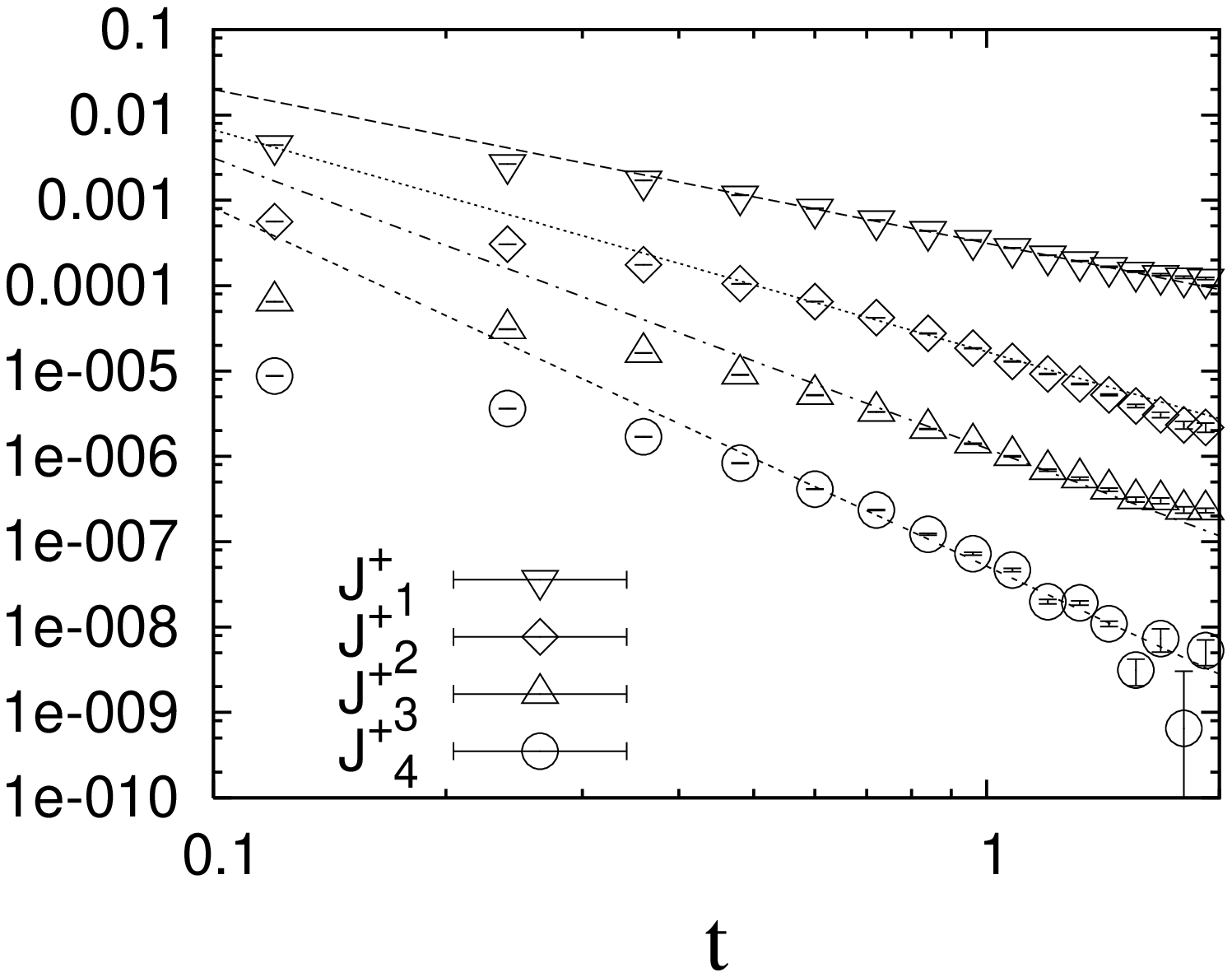,%
angle=270,width=7.4cm}
\caption{The log-log plot of the correlator 
$\Bigl \langle J^+_{\ell}(t)\, J^+_{\ell}(0) \Bigr\rangle$
with  $\ell=1,3$ for $N=2$ (Left) and 
with  $\ell=1,2,3,4$ for $N=3$ (Right).
The cutoff parameters are chosen as $\beta=4$ and $\Lambda=16$.
The straight lines represent the power-law behavior predicted
by the gauge-gravity correspondence.
}
\label{fig:Jp_N3C16T025}
}

The existence of the undetermined analytic terms 
in (\ref{correlator_momentum_space})
makes the extraction of $\nu$ more difficult, in particular,
for large $\nu$.
This problem can be avoided if one 
can make the inverse Fourier transformation
(\ref{inverse-Fourier-corr}) numerically.
The analytic terms are transformed into local terms,
and hence do not affect the power-law behavior (\ref{sugraresult})
of correlation functions in the real space,
from which one can extract the exponent $\nu$.
However, there are cases in which
the inverse Fourier transformation is not possible numerically.
This can happen either due to the IR behavior or due to
the UV behavior.
Correlation functions for operators with $\nu < -\frac{1}{2}$
(\emph{e.g.}, $T^{++}_2$) behave as $\sim |p|^{2\nu}$  at small $p$,
and hence the inverse Fourier transform is divergent at $p\sim 0$.
The correlation function for operators (\emph{e.g.}, $T^+_\ell$)
including a derivative does not fall off in the momentum space
at large $p$, which makes the inverse Fourier transformation
numerically unstable.
Even in these cases, it turns out that 
the results we obtain directly in the momentum space
are in good agreement with the prediction 
from the gauge-gravity correspondence.

%

In what follows we present our results for
$J^{+}_\ell$, $T^{++}_\ell$ and $T^+_\ell$ in order.
In the $N=2$ case the correlation function becomes
identically zero for 
$J^{+}_\ell$ ($\ell  : \, $even), $T^{++}_\ell$ 
($\ell  : \, $odd) and
$T^+_\ell$ ($\ell  : \, $even)
due to properties of the Pauli matrices,
and hence we omit these cases.



Let us start with the correlation functions of
operators $J_{\ell}^+$,
for which the inverse Fourier transform can be calculated numerically.
If we naively make the inverse transform, however,
the correlation function in the real space
shows oscillating behavior with the period 
$\delta t \sim \beta/(2\pi \Lambda)$.
This is well-known as the Gibbs phenomenon,
and it is an artifact of the sharp cutoff in the momentum space.
In the present case, we may naturally expect that
the large-$p$ behavior of the two-point correlation functions is given by
\begin{eqnarray}
\left\langle\tilde{J}_{\ell}^+(p) \, \tilde{J}_{\ell}^+(-p)\right\rangle
\sim \frac{\kappa}{p^2} \ .
\label{ansaz:extrapolation in momentum space}
\end{eqnarray}
Indeed our data can be nicely fitted to
this behavior at large $p$,
and we can obtain the coefficient $\kappa$ reliably.
The correlation function in the real space can then be obtained as
\begin{eqnarray}
\Bigl \langle J_{\ell}^+(t)\, J_{\ell}^+(0) \Bigr\rangle
=
\left\langle\tilde{J}_{\ell 0}^+  \, 
\tilde{J}_{\ell 0 }^+ \right\rangle
+
\sum_{n>0}
2\cos\left( \frac{2 \pi n t}{\beta}\right)
\left\langle\tilde{J}_{\ell n}^+ \, \tilde{J}_{\ell , -n }^+ \right\rangle \ ,
\label{inv-Fourier-tr}
\end{eqnarray} 
where we extend the sum over $n$ beyond $\Lambda$ using
the form (\ref{ansaz:extrapolation in momentum space})
up to $n=1000$.
The results obtained in this way are shown in fig.~\ref{fig:Jp_N3C16T025}
for $N=2$ (Left) and $N=3$ (Right).
The cutoff parameters are $\beta=4$ and $\Lambda=16$.
We fit our data points for $n=10,11$ and $12$ to
(\ref{ansaz:extrapolation in momentum space}),
and use the form for $12 < n \le 1000$ in the sum (\ref{inv-Fourier-tr}).
Straight lines represent the power-law behavior ${\rm const.}/t^{2\nu+1}$
with $\nu=2\ell /5$, which is predicted for $N=\infty$
from the gauge-gravity correspondence.
In the large-$t$ region ($t \gtrsim 0.5$), the power law is reproduced 
with remarkable precision up to 
the IR cutoff scale ($t \lesssim \beta/2$).
Moreover, the results for $N=2$, which are obtained only for odd $\ell$
for the reason mentioned above, 
already show the power-law behavior with the same exponent 
as predicted by the supergravity analysis.
This suggests that the exponents are actually
independent of $N (\ge 2)$. 

    \FIGURE[t]{
    \epsfig{file=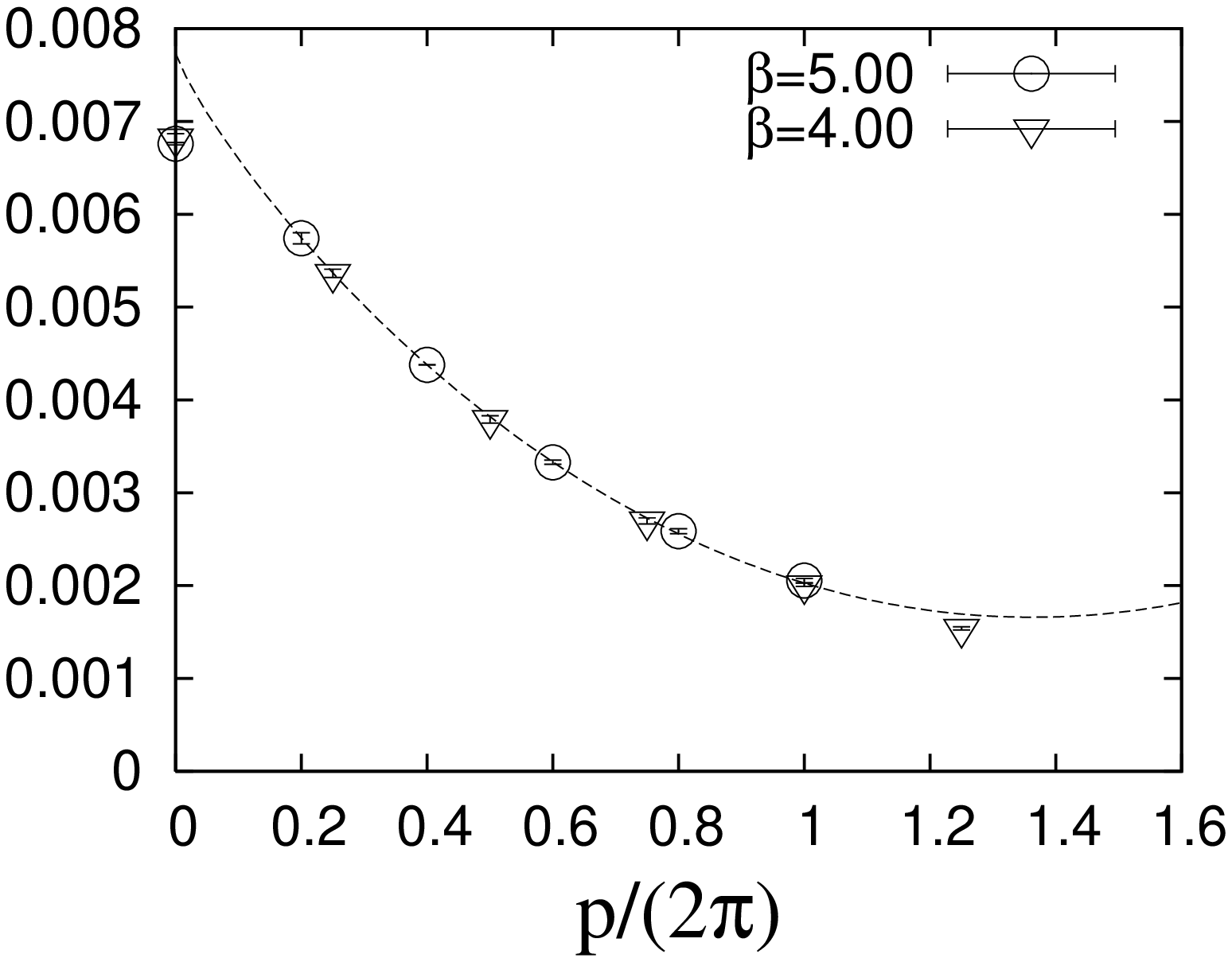,%
angle=270,width=7.4cm}
    \epsfig{file=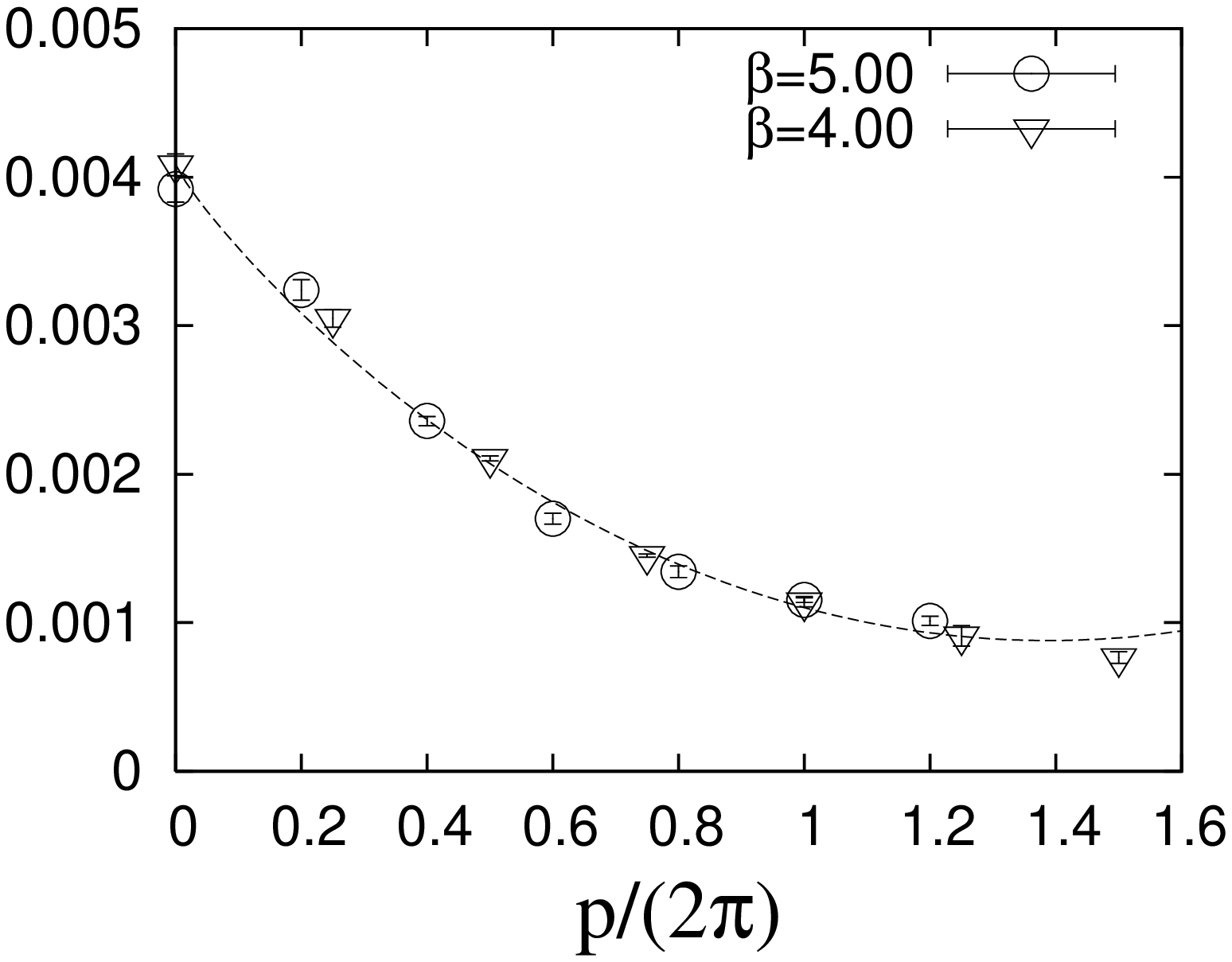,%
angle=270,width=7.4cm}
\caption{The correlator 
$\Bigl\langle \tilde{J}^+_1(p) \, \tilde{J}^+_1(-p) \Bigr\rangle$ 
is plotted as a function of $p$
for $N=2$ (Left) and $N=3$ (Right).
The extrapolation to $\Lambda=\infty$ is made at each $p$.
The dashed line represents a fit to the behavior
(\ref{J^+_1 mom space}),
from which we obtain $\nu=0.42(6)$ for $N=2$
and $\nu=0.43(5)$ for $N=3$.
The predicted value is $\nu = 2/5$.
}
\label{fig:Jp2_MomemtumSpace_mono_v1}
}


The consistency with the gauge-gravity correspondence
can also be seen directly in the momentum space.
Since we are interested in the IR behavior, 
let us restrict ourselves to the behavior at small $p$. 
According to the gauge-gravity correspondence, 
the first few terms relevant at small $p$ for 
$\ell =1,2,3,4$ are
\begin{equation}
\left\langle\tilde{J}^{+}_\ell(p)\, \tilde{J}^{+}_\ell(-p)\right\rangle
 \simeq
\left\{
\begin{array}{lcl}
a+b\, |p|^{2\nu}+c \, p^2 & \mbox{~~~~~}  &\mbox{for~}\ell=1,2  \ , \\
a+b \, p^2+c \, |p|^{2\nu} & \mbox{~~~~~}  &\mbox{for~}\ell=3,4  \ , 
\end{array}
\right.
\label{J^+_1 mom space}
\end{equation}
where $\nu= 2 \ell/5$.
In fig.~\ref{fig:Jp2_MomemtumSpace_mono_v1}
we show our results for $J^+_1$ with $N=2$ (Left) and $N=3$ (Right).
At each $p$ we made an extrapolation to $\Lambda=\infty$
assuming that
the finite $\Lambda$ effect is of the order of 
${\rm O}(1/\Lambda)$.\footnote{In most cases we used the data for 
$\Lambda=8,12,16$. In the $N=3$ case with
$p/(2\pi) \le 3/\beta$ for $\beta\ge 5$,
we had to use the data for $\Lambda=6,8,12$
since the data for $\Lambda=16$ had too large statistical errors.
The extrapolation to $\Lambda=\infty$ seems to be fine, though,
since finite $\Lambda$ effects are less severe in the small $p$ region.
}
By fitting our results at 
$0 < p/(2\pi)\le 1$
to the behavior (\ref{J^+_1 mom space}),
we obtain $\nu=0.42(6)$ and $\nu=0.43(5)$ for $N=2,3$, respectively,
which are consistent with the predicted value $\nu=2/5$. 
%
The same analysis can be performed also for $J^+_\ell$
with $\ell=2,3,4$.
The values of the parameters obtained by fitting our data for $N=3$
are summarized in Table \ref{tab:tableJ}.
The fits are reasonable, and they
are consistent 
with the predictions from the gauge-gravity correspondence.
%
%

\TABULAR[t]{|c||c|c|c|c||c|} {
\hline
&  $a$ & $b$ & $c$ & $\nu$ & $\nu_{{\rm pred}}$
 \\ \hline\hline
$J_1^+$  &  $0.40(1) \times 10^{-2}$ 
& $-0.8(1) \times 10^{-3}$  &
$0.32(8) \times 10^{-4}$
& $0.43(5)$ & $0.4$ \\ \hline
$J_2^+$ & $0.438(1) \times 10^{-3}$ 
& $-0.58(7) \times 10^{-4}$ 
& $0.26(9) \times 10^{-4}$ & $0.84(3)$ & $0.8$ \\ \hline
%
%
$J_3^+$ & $0.47\times 10^{-4}$ 
& $-0.38\times 10^{-5}$ &
$0.17\times 10^{-5}$ & 1.18 & $1.2$ \\ \hline
$J_4^+$ & $0.582(6) \times 10^{-5}$ 
& $-0.12(5) \times 10^{-6}$ &
$0.57(4) \times 10^{-8}$ & $\ast$ & $1.6$ \\
\hline
}
{The parameters obtained by fitting our results
for the correlators  $\Bigl\langle \tilde{J}^+_\ell(p) \, 
\tilde{J}^+_\ell(-p)\Bigr\rangle$ 
with $N=3$ to the behavior (\ref{J^+_1 mom space}).
The values of $\nu$ predicted from the gauge-gravity correspondence
are shown on the right most column.
For $J_3^+$, the fitting errors are not shown 
since we can use only 4 data points for the fit.
For $J_4^+$, we present the values of $a$, $b$ and $c$
obtained by fixing $\nu$ to the predicted value $1.6$
since the fitting errors turned out to be large otherwise.
\label{tab:tableJ}}



    \FIGURE[t]{
    \epsfig{file=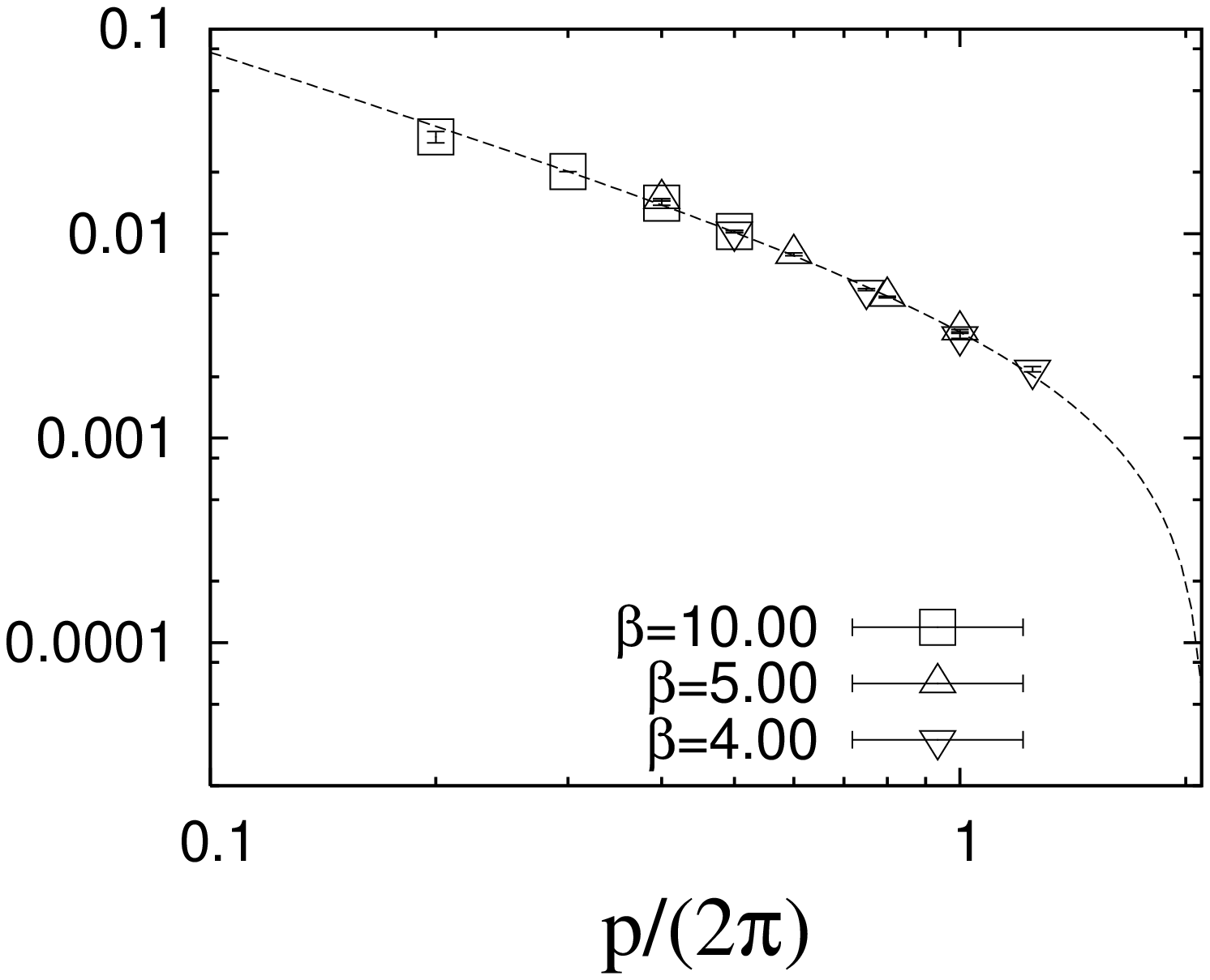,%
angle=270,width=7.4cm}
    \epsfig{file=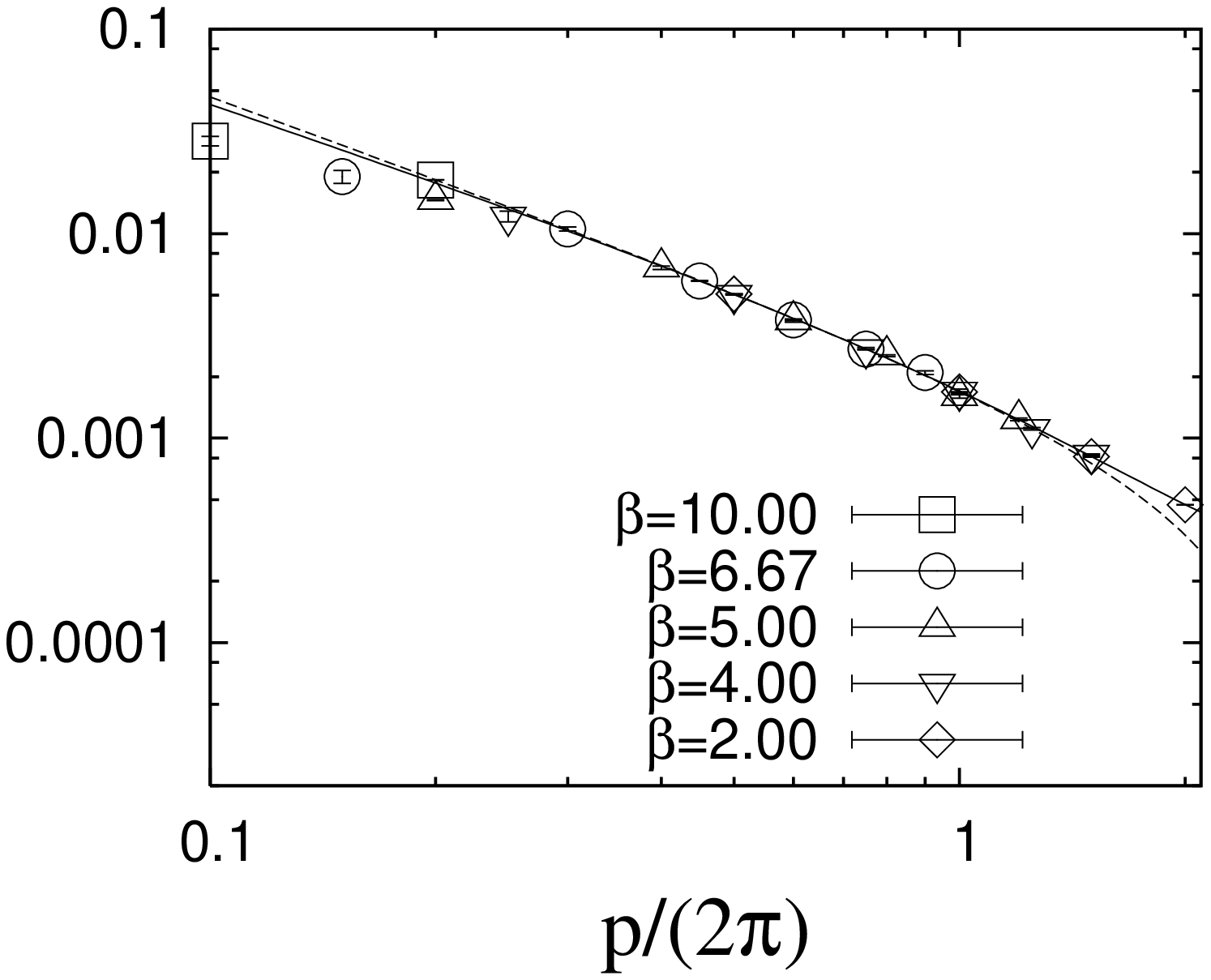,%
angle=270,width=7.4cm}
\caption{The log-log plot of the correlator
$\left\langle\tilde{T}^{++}_2(p) \, \tilde{T}^{++}_2(-p)
\right\rangle$ for $N=2$ (Left) and $N=3$ (Right). 
The extrapolation to $\Lambda=\infty$ is made at each $p$.
The dashed line is a fit
to the behavior (\ref{asmp-p2}),
from which we obtain $\nu=-0.57(3)$
and $\nu=-0.66(2)$ for $N=2$ and $N=3$, respectively.
For the $N=3$ case, we can fit the data up to $p/(2\pi)=2$ 
as shown by the solid line
by including a higher order term $d \, |p|^{2\nu+2}$ in (\ref{asmp-p2}), 
which gives
$\nu = - 0.61(2)$.
The predicted value is $\nu = -3/5$.
%
%
}
\label{fig:T^++_2_SU2}
}

    \FIGURE[h]{
    \epsfig{file=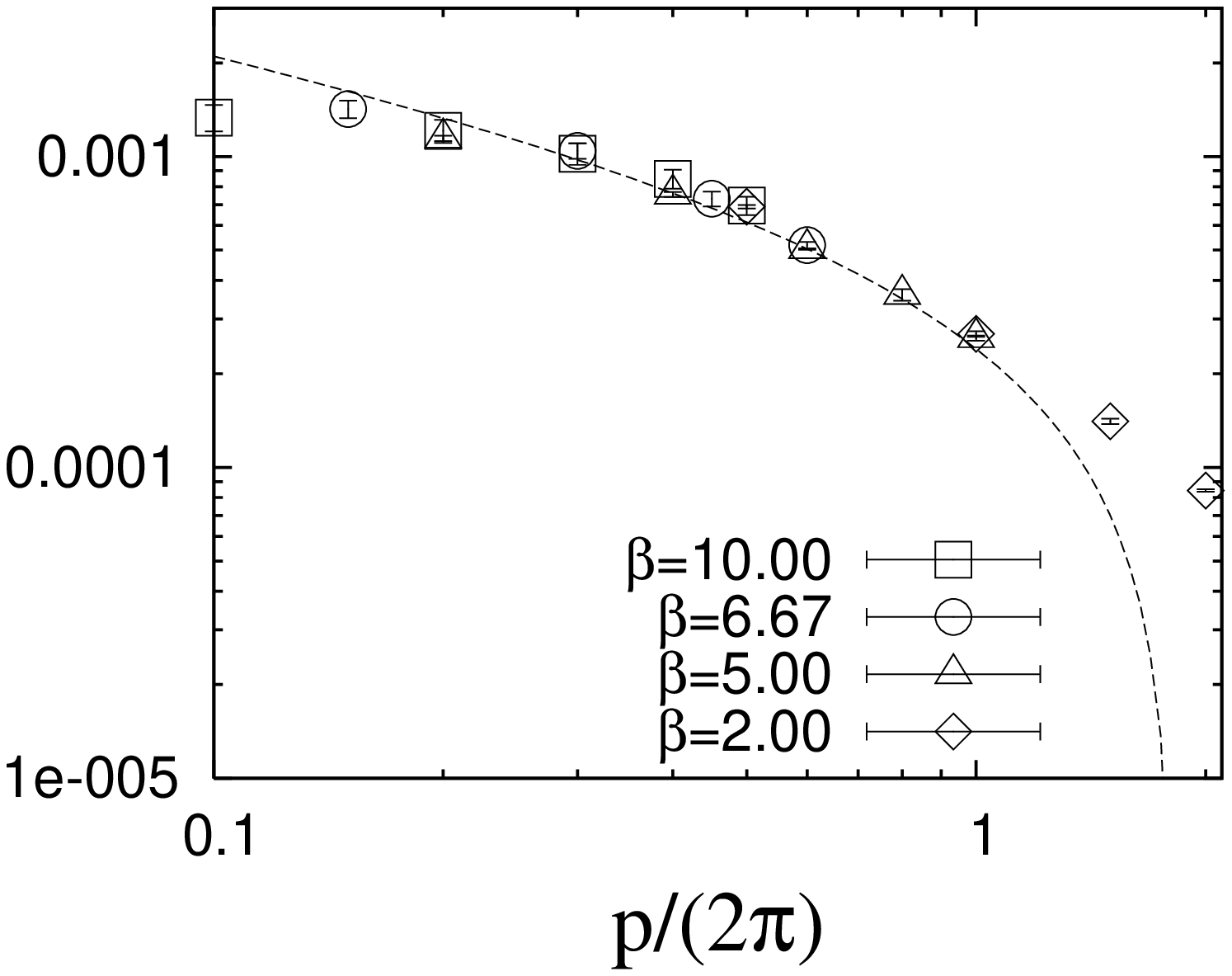,%
angle=270,width=9cm}
\caption{The log-log plot of the correlator 
$\left\langle \tilde{T}^{++}_3(p) \, 
\tilde{T}^{++}_3(-p) \right\rangle$ for $N=3$.  
The extrapolation to $\Lambda=\infty$ is made at each $p$.
The dashed line represents a fit
to the behavior (\ref{asmp-p2})
within the range is $2/\beta \le p/(2\pi) \le 0.8$,
which gives $\nu=-0.2(1)$. 
The predicted value is $\nu=-1/5$.
}
\label{fig:Tpp3_N3}
}

    \FIGURE[t]{
    \epsfig{file=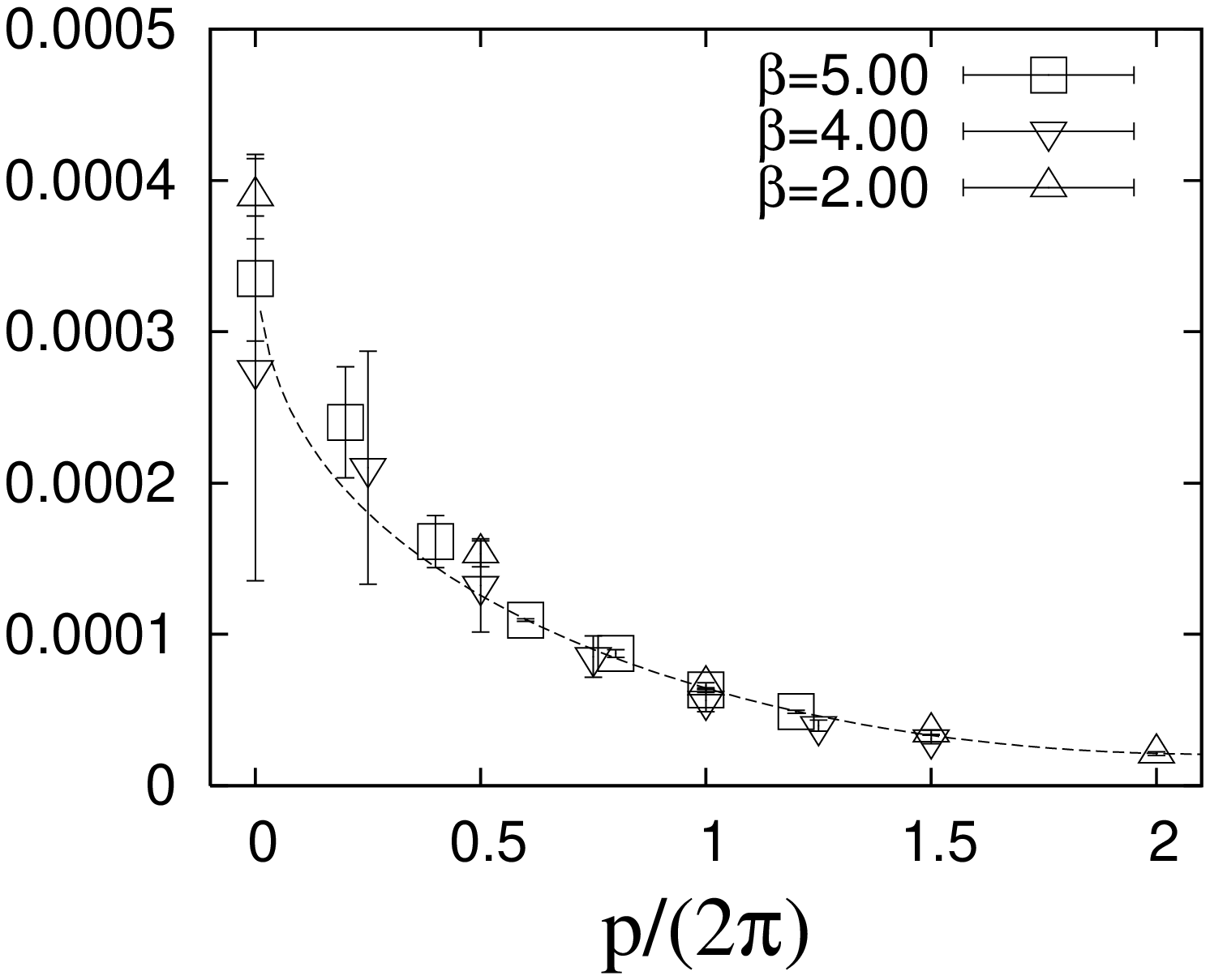,%
angle=270,width=7.4cm}
    \epsfig{file=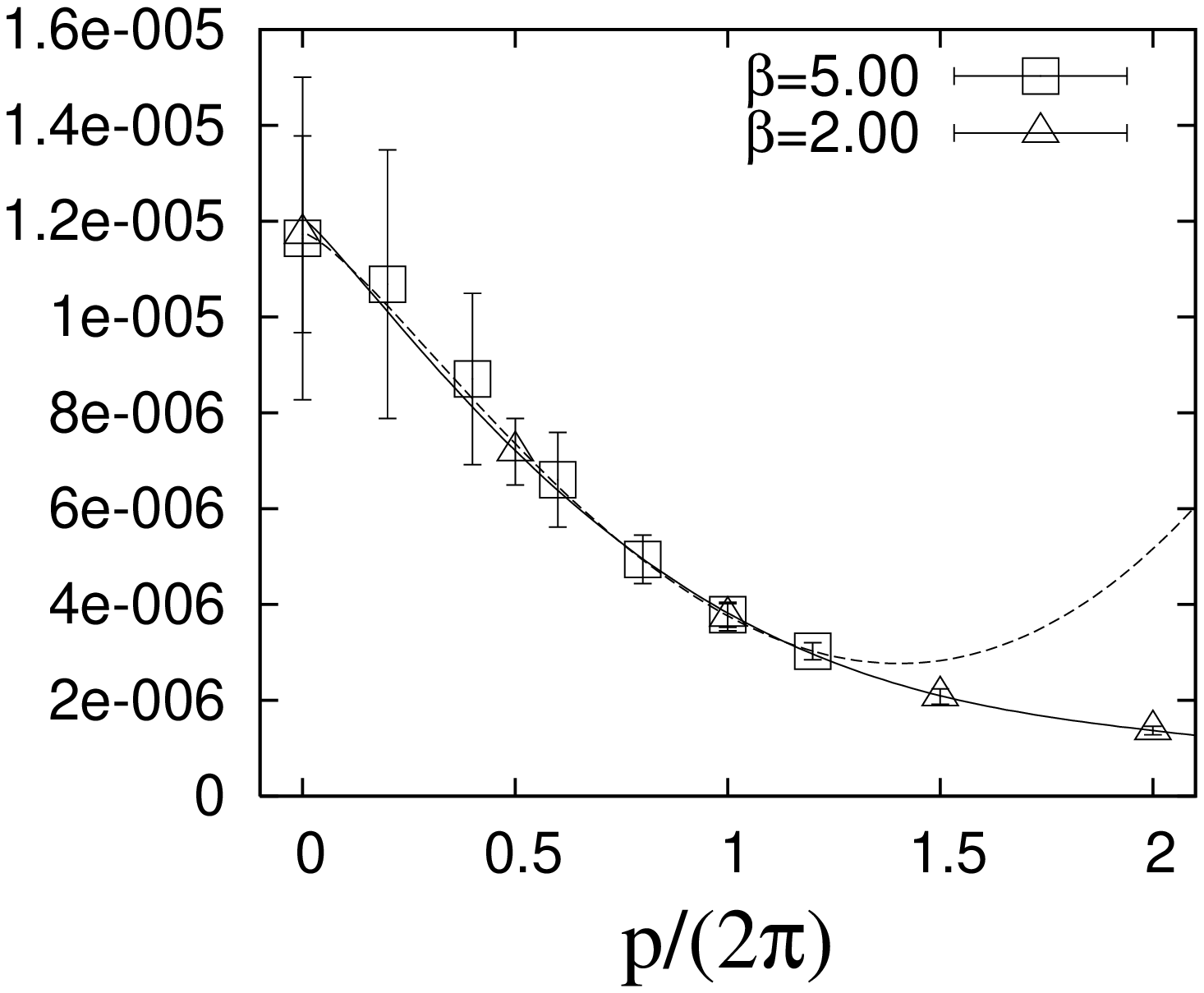,%
angle=270,width=7.4cm}
\caption{(Left) The correlator
$\left\langle \tilde{T}^{++}_\ell(p) \, \tilde{T}^{++}_\ell(-p)
\right\rangle$ 
is plotted as a function of $p$ 
for $\ell=4$ (Left) and $\ell=5$ (Right)
in the $N=3$ case.
The extrapolation to $\Lambda=\infty$ is made at each $p$.
The dashed line represents a fit to the behavior (\ref{predTpp45}),
which gives 
$\nu=0.19(3)$ and $\nu=0.7(1)$
for $\ell=4$ and $\ell=5$, respectively.
For the $\ell=5$ case, we can fit the data up to $p/(2\pi)=2$ 
as shown by the solid line
by including a higher order term $d \, |p|^{2\nu+2}$ in (\ref{predTpp45}), 
which gives 
$\nu=0.60(9)$.
The predicted values are 
$\nu=1/5$ and $\nu=3/5$
for $\ell=4$ and $\ell=5$, respectively.
}
\label{fig:T^++_4_SU3}
}

Let us next consider the operator $T_{\ell}^{++}$ 
for $\ell=2,3,4,5$.
The gauge-gravity correspondence predicts
$\nu=(2\ell-7)/5$ \cite{SY}.
The $\ell=2,3$ cases are of particular interest
since the supergravity analysis predicts
$\nu=-3/5$ and $\nu=-1/5$, respectively, which are negative.
From (\ref{correlator_momentum_space}), 
the correlation function in the momentum space is 
expected to behave as
\begin{alignat}{3}
\left\langle\tilde{T}^{++}_\ell(p) \, \tilde{T}^{++}_\ell(-p)\right\rangle
&=
a \, |p|^{2\nu} + b  \quad  \quad \mbox{for $\ell=2,3$}
\label{asmp-p2}
\end{alignat}
at small $p$.
Note that the leading term is divergent 
in the IR limit $p \rightarrow 0$ unlike
the previous cases.
In particular, the inverse Fourier transform
(\ref{inverse-Fourier-corr})
is ill-defined for $\ell=2$.
This is somewhat reminiscent of the free-field behavior 
$\sim 1/p^2$ albeit with a milder form.
In fig.~\ref{fig:T^++_2_SU2} 
we show a log-log plot in the case of $\ell=2$
for $N=2$ (Left) and $N=3$ (Right).
%
For the $N=2$ case, we can fit all the data in the figure 
with $p/(2\pi) \ge 1/\beta$ to (\ref{asmp-p2}), which gives
$\nu = -0.57(3)$.
For the $N=3$ case, we can fit the data within $1/\beta < p/(2\pi) \le 1$
as represented by the dashed line, which gives $\nu=-0.66(2)$.
By including a higher order term $d \,  |p|^{2\nu+2}$ in (\ref{asmp-p2}), 
however, we can fit all the data in the figure with $p/(2\pi) \ge 1/\beta$
as represented by the solid line,
which gives $\nu = - 0.61(2)$.
These results are consistent with the predicted value $\nu=-3/5$.
%
The deviation from the behavior (\ref{asmp-p2})
at smaller $p$ around the IR bound ($\sim \beta^{-1}$) 
is naturally interpreted as finite $\beta$ effects. 
%
In fig.~\ref{fig:Tpp3_N3} we plot the result for $T_3^{++}$
in the $N=3$ case. 
Fitting our data to the behavior (\ref{asmp-p2}),
we obtain $\nu=-0.2(1)$, which is consistent with
the predicted value $\nu=-1/5$.

\TABULAR[t]{|c||c|c|c|c||c|} {
\hline
& $a$ & $b$ & $c$ & 
$\nu$ & $\nu_{{\rm pred}}$ \\ \hline\hline
$T_2^{++}$ & $ 0.249(6) \times 10^{-1}$ 
& $ -0.13(2) \times 10^{-2}$ 
& $\ast$  &
$-0.61(2)$ & $-0.6$ \\ \hline
%
$T_3^{++}$ & $0.23(5) \times 10^{-2}$
 & $-0.8(6) \times 10^{-3}$ & $\ast$ & 
$-0.2(1)$ & $-0.2$ \\ \hline
$T_4^{++}$ & $0.38(3) \times 10^{-3}$ 
& $-0.16(3) \times 10^{-3}$ 
& $0.45(7) \times 10^{-6}$ 
& $0.19(3)$ & $0.2$ \\ \hline 
$T_5^{++}$ & $0.12(5)\times 10^{-4}$ 
& $-0.17(3) \times 10^{-5}$ 
& $0.2(1) \times 10^{-6}$ 
& $0.60(9)$ & $0.6$ \\
%
%
\hline
}
{The parameters obtained by fitting our results for the correlators
$\left\langle\tilde{T}^{++}_\ell(p) \, \tilde{T}^{++}_\ell(-p) \right\rangle$ 
with $N=3$ to (\ref{asmp-p2}) for $\ell=2,3$ and
to (\ref{predTpp45}) for $\ell=4,5$.
The values of $\nu$ predicted from the gauge-gravity correspondence
are shown on the right most column.
For $\ell = 2$ and $5$,
we also include a term $d \,  |p|^{2\nu+2}$
with the coefficient 
$d=  0.95(6) \times 10 ^{-4}$ and
$d= -0.24(7)\times 10^{-8}$, respectively.
\label{tableT}}


In the $\ell=4,5$ cases, the gauge-gravity correspondence predicts
\beq
\left\langle\tilde{T}^{++}_{\ell}(p) \, 
\tilde{T}^{++}_{\ell}(-p)\right\rangle
= a+b \, |p|^{2\nu} +c \, p^2
\label{predTpp45}
\eeq 
at small $p$, where $\nu=1/5,3/5$ for $\ell=4,5$, respectively.
In fig.~\ref{fig:T^++_4_SU3} we plot the 
correlation functions for $T^{++}_{4}$ (Left) and $T^{++}_{5}$ (Right)
with $N=3$. 
The data at small $p$ have large statistical errors.
We consider this as a consequence of the fact
that the operators $T^{++}_{\ell}$
with large $\ell$ are composed of large powers of $X$ without
derivatives or commutators, and hence they are more affected by
the large fluctuations of the low-momentum modes of the $X$ field.
%
For the $N=2$ case, we can fit all the data 
in the figure to (\ref{predTpp45}), which gives
$\nu=0.19(3)$. 
For the $N=3$ case, we can fit the data with $p/(2\pi) \le 1.2$
as represented by the dashed line, which gives
$\nu=0.7(1)$.
By including a higher order term 
$d \,  |p|^{2\nu+2}$ in (\ref{predTpp45}), however,
we can fit all the data in the figure 
as represented by the solid line,
which gives 
$\nu=0.60(9)$.
These results are consistent with the predicted values
$\nu=1/5,3/5$ for $\ell=4,5$, respectively.
In Table \ref{tableT}
we present the values of the fitting parameters obtained
for $T^{++}_{\ell}$ ($\ell=2,3,4,5$)
in the $N=3$ case.



As the last example of operators corresponding to supergravity modes,
let us consider the operator $T_\ell^{+}$ defined by (\ref{Tplus}).
The peculiarity of this operator is that it involves a derivative
with respect to $t$.
As a result, the corresponding two-point correlation functions
do not decrease as $\sim 1/p^2$ unlike the previous examples.
Although finite $\Lambda$ effects turned out to be rather large,
we were able to extrapolate our results obtained
at $\Lambda=6,8,12,16$ to $\Lambda=\infty$.
In fig.~\ref{T^+_2} we plot
$\left\langle \tilde{T}^{+}_2(p) \, \tilde{T}^{+}_2(-p)\right\rangle$ 
for $N=3$, $\beta=5$. Fitting the data 
in the region $0 \le  p/(2\pi) < 1$ to the behavior
\begin{eqnarray}
\left\langle\tilde{T}^{+}_2(p) \, \tilde{T}^{+}_2(-p)\right\rangle
= a+b \, |p|^{2\nu} + c \, p^2 \ ,
\label{T^+_2 mom space} 
\end{eqnarray} 
we obtain $\nu= 0.80(3)$, which is 
consistent with the predicted value $\nu=4/5$.  

    \FIGURE[t]{
    \epsfig{file=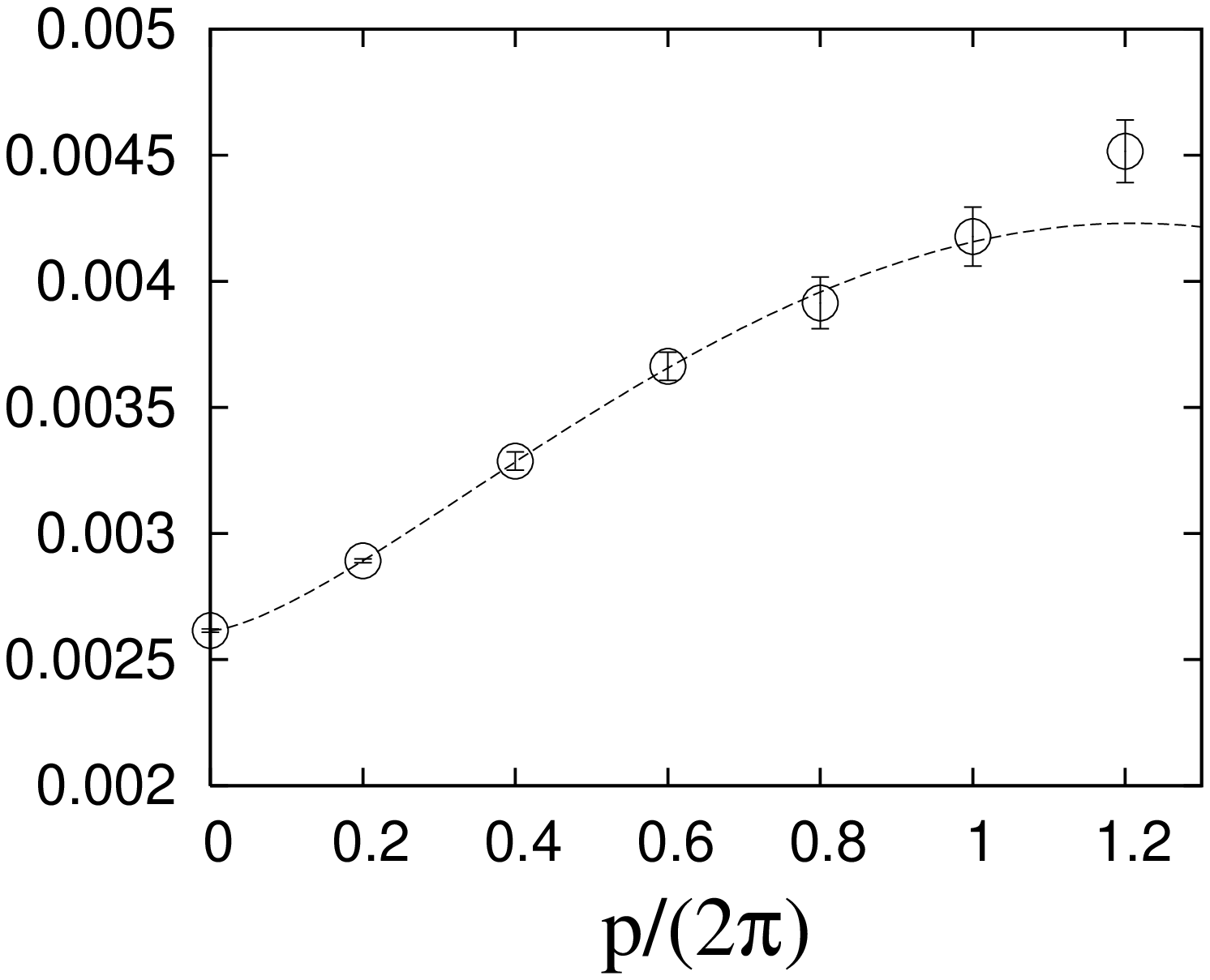,%
angle=270,width=9cm}
\caption{The correlator
$\left\langle\tilde{T}^+_2(p) \, \tilde{T}^+_2(-p)
\right\rangle$ 
is plotted for $N=3$ and $\beta=5$. 
The extrapolation to $\Lambda=\infty$ is made at each $p$.
The dashed line represents a fit to the behavior
(\ref{T^+_2 mom space})
with $a=2.614(2)\times 10^{-3}$, 
$b=3.1(3)\times 10^{-4}$,
$c=-1.1(3)\times 10^{-4}$, 
and $\nu = 0.80(3)$.
The value of $\nu$ predicted from the gauge-gravity correspondence
is $\nu=4/5$.
}
\label{T^+_2}
}

Finally, let us briefly comment on the sign of the leading term
in 
all the correlation functions studied in this paper.
Our data together with the formula \eqref{ftransform} 
shows that 
the correlation functions obey
the positive IR behavior in the real space
suggested from the (reflection) positivity,
except for the operators $T^{++}_{\ell}, T^{+}_{\ell}$ with $\ell=2$.
%
In the case of $T^{++}_2$, we note that
the naive L$^1$ integrability condition
for the inverse Fourier transformation 
is violated in the IR region since
$\left\langle\tilde{T}^{++}_2(p) \, \tilde{T}^{++}_2(-p)\right\rangle
\sim |p|^{-6/5}$ at $p\rightarrow 0$.
The non-positivity may 
be attributed to this property
in analogy with 
the apparent violation of positivity
that occurs 
for the free-field propagator  $\Bigl\langle X(t)X(0)\Bigr\rangle 
=-|t|/2 + {\rm const.}$, which is related with 
the divergent momentum-space behavior $1/p^2$ at $p=0$. 
In the case of $T^{+}_2$,
the momentum-space correlation function 
does not decrease for $p\rightarrow \infty$ as exhibited in fig.~\ref{T^+_2},
and hence the UV behavior does not satisfy the L$^1$ integrability, either.
Note, however, that the non-integrability at the UV region 
does not directly affect the leading IR behavior. 
Therefore the situation is rather unclear.

\subsection{results for stringy excited modes}
\label{sec:res-stringy}

In this subsection we present our Monte Carlo results for 
operators corresponding to stringy excited modes. 
Let us first consider the BMN-type operator (\ref{typ-ope}).
The result (\ref{stringamplitude})
based the gauge-gravity correspondence
suggests that the corresponding two-point functions behave as
\begin{eqnarray}
\Bigl\langle {\cal O}_{ij,n}^J(t) \, {\cal O}_{ij,n}^{J\dagger}(0)
\Bigr\rangle
\sim 
a\exp\left(-b \, t^{3/5}\right) \  ,
\label{bmnprediction}
\end{eqnarray} 
up to a power behaved correction factor.
%
This prediction was obtained in the plane-wave limit,
which is justified when the angular momentum $J$ and 
the wave number $n$ are both large.
%

In what follows we calculate the
correlation function on the gauge theory side
and compare the results with the prediction (\ref{bmnprediction}).
For that purpose we make an inverse Fourier transformation
as we did for the operator $J^{+}_\ell$ in the previous section.
In order to avoid the Gibbs phenomenon associated with the sharp
UV cutoff, we extrapolate our data to larger momenta
assuming the form $\frac{\kappa}{p^2+m^2}$.
We take the UV cutoff to be $\Lambda=16$, 
and determine the parameters 
$\kappa$ and $m^2$ by fitting the data at $p=2\pi k/\beta$ for 
$k=8,\cdots,12$.
Similarly to the case of $T^{++}_\ell$ with $\ell=4,5$
shown in fig.~\ref{fig:T^++_4_SU3}, 
the two-point functions of the BMN-type operators 
for $J\ge 4$ have large statistical errors
in the small $p$ region, which propagates to the real-space
correlator through the inverse Fourier transformation.
Therefore, we restrict ourselves to $J=2,3$.

    \FIGURE[t]{
    \epsfig{file=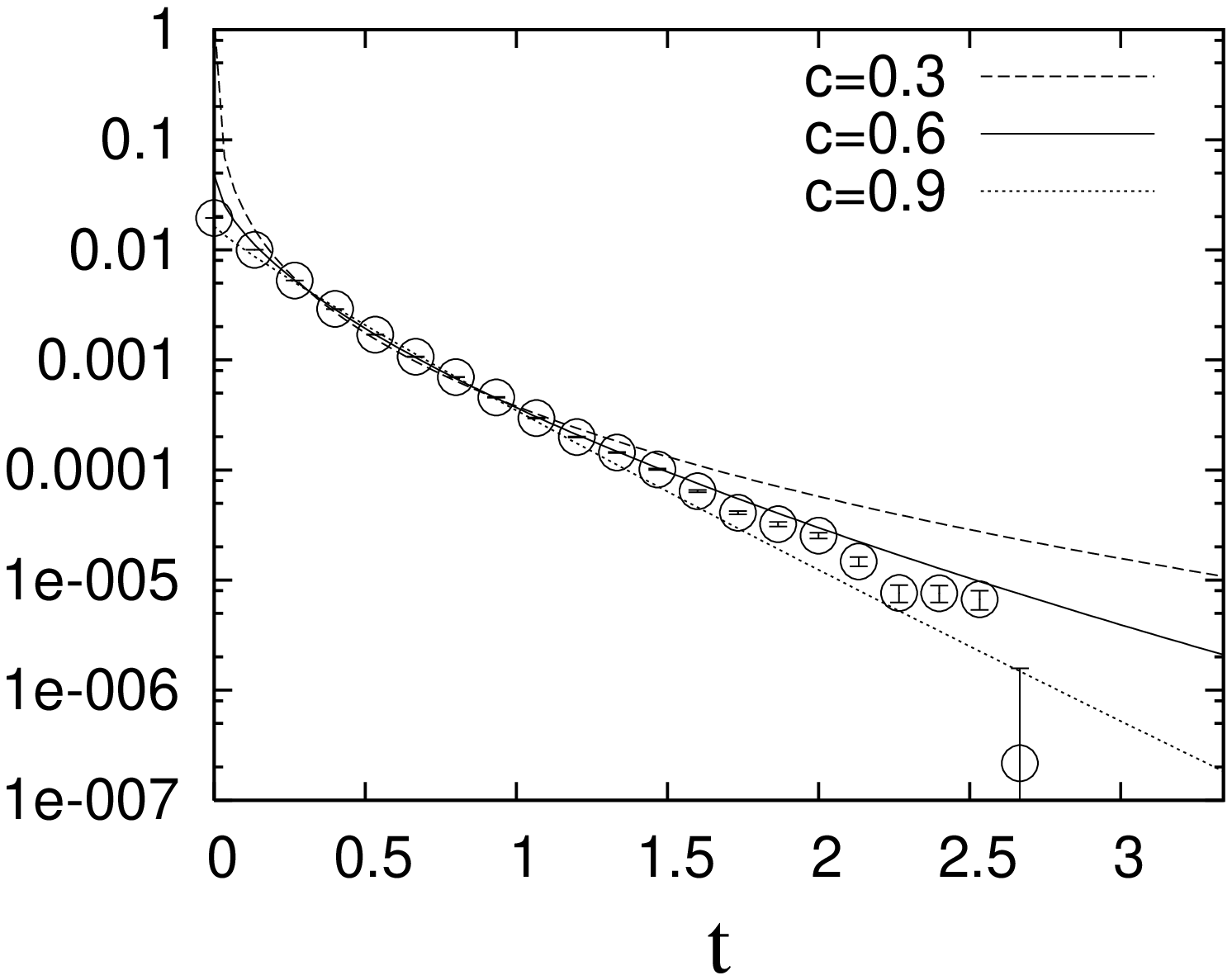,%
angle=270,width=7.4cm}
    \epsfig{file=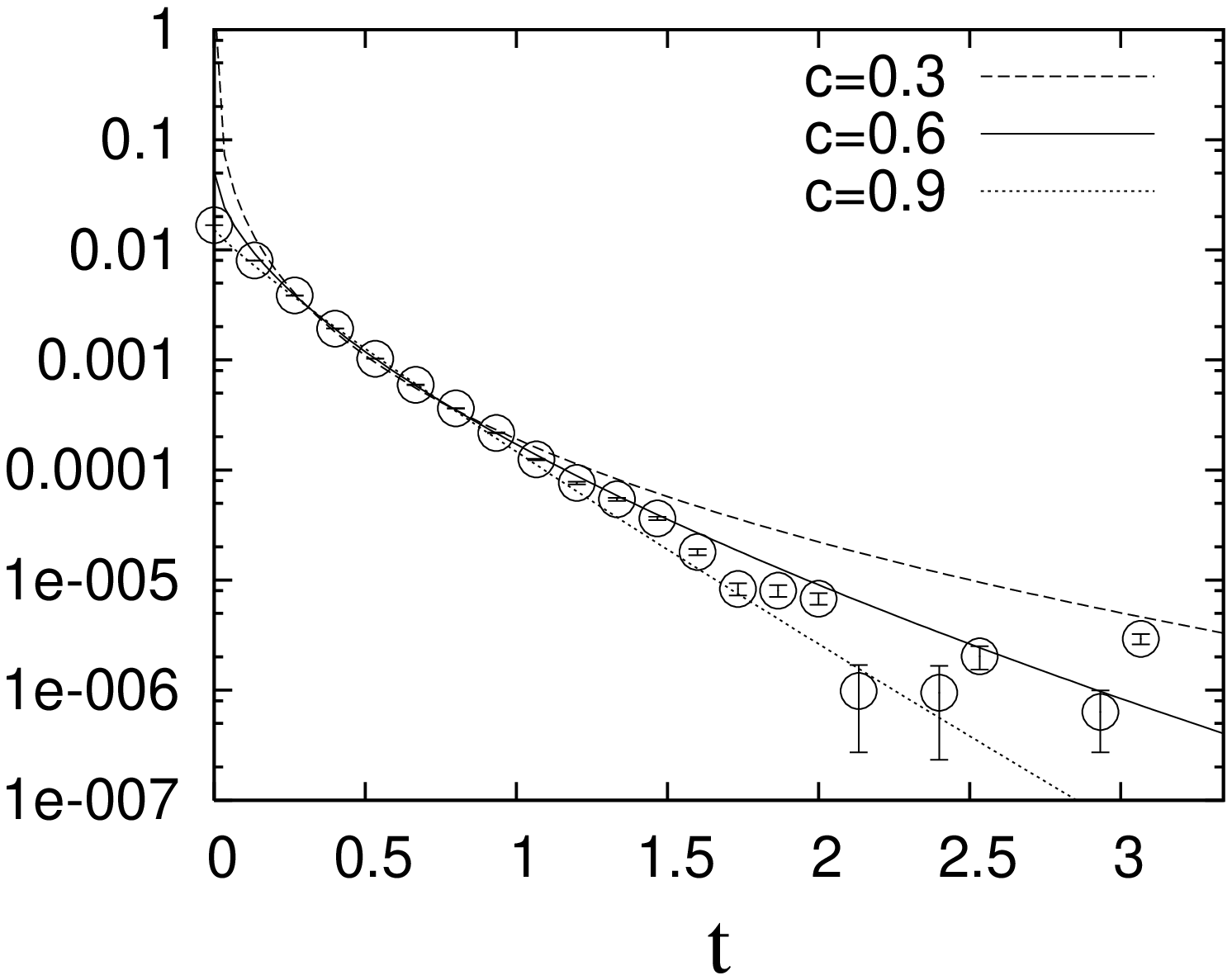,%
angle=270,width=7.4cm}
\caption{The two-point correlation function of
the BMN-type operator (\ref{typ-ope})
is plotted in the real space
for $J=2$ (Left) and $J=3$ (Right) with $n=1$.
The matrix size is $N=3$ and the cutoff parameters
are $1/\beta=0.15$ and $\Lambda=16$. 
The curves represent the fits to the form 
$a\exp\left(-b \, t^{c}\right)$ with $c=0.3, 0.6, 0.9$. 
Fitting the data in $0.2\le t\le 1.5$, we obtain
$c=0.63(1)$ and $c=0.66(2)$ for $J=2$ and $J=3$, respectively.
}
\label{fig:BMN_J2n1_T015}
}

Figure~\ref{fig:BMN_J2n1_T015} 
shows the results for BMN-type operators 
with $J=2$ and $J=3$, respectively.
The wave number $n$ in the operator (\ref{typ-ope}) is set to $n=1$.
Fitting the data within $0.2\le t\le 1.5$ to the form 
$a\exp\left(-b \, t^{c}\right)$,
we obtain $c=0.63(1)$ and $c=0.66(2)$
for $J=2$ and $J=3$, respectively,
which are close to the value $3/5$ predicted for large $J$ and $n$.
The deviations observed at large $t \, (\gtrsim 2) $ can
be attributed to finite-$\beta$ effects.
The coefficient $b$ is predicted 
to be proportional to $n/J$ for large $n$ and $J$ with $n \gg J$.
The values of $b$ obtained 
from the fit by fixing $c$ to be $3/5$
are $b=4.70$ and $b=5.40$ for $J=2$ and $J=3$, respectively.
This suggests that the coefficient $b$ has more nontrivial
dependence than $b \propto n/J$ for small $n$ and $J$.


    \FIGURE[h]{
    \epsfig{file=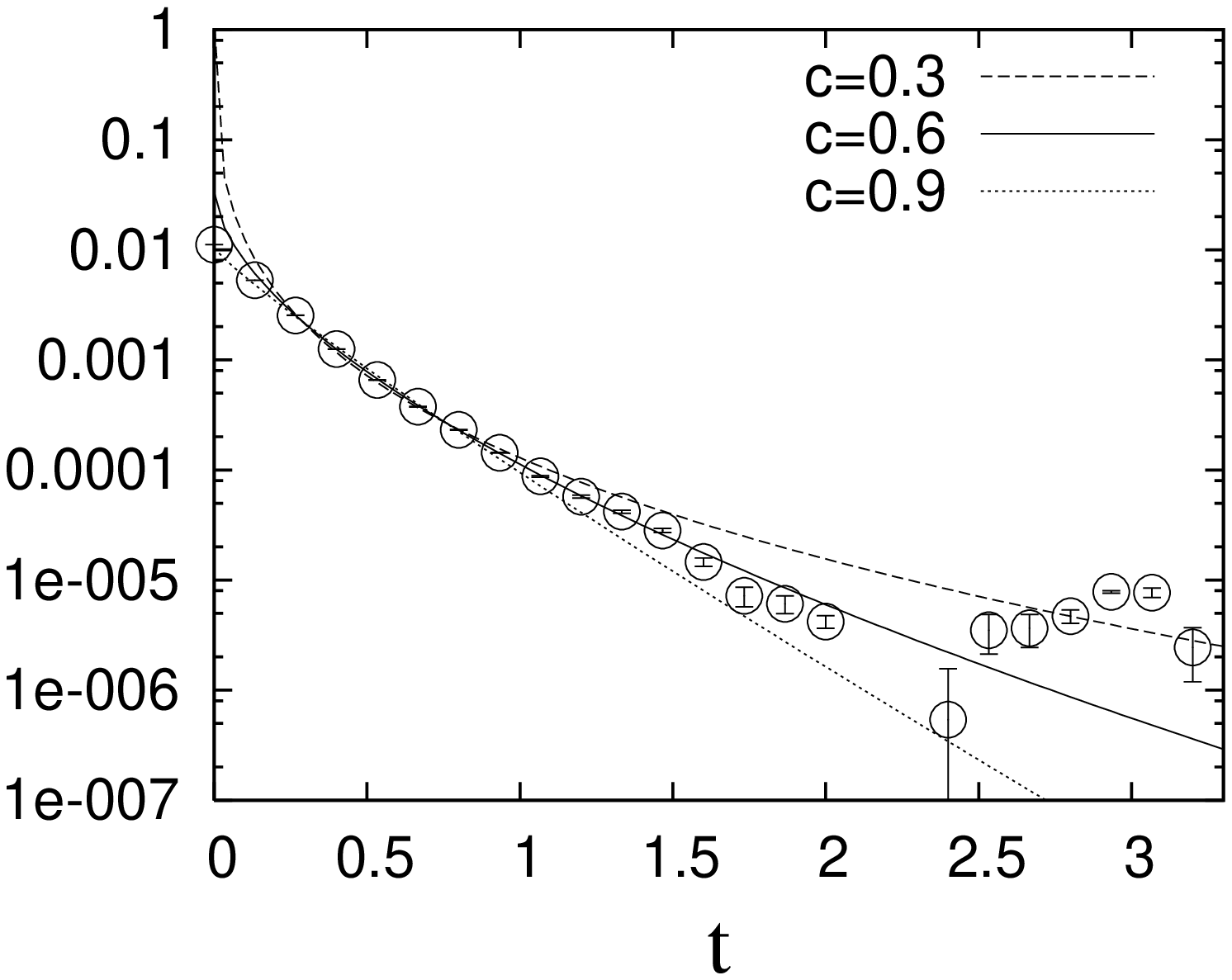,%
angle=270,width=7.4cm}
    \epsfig{file=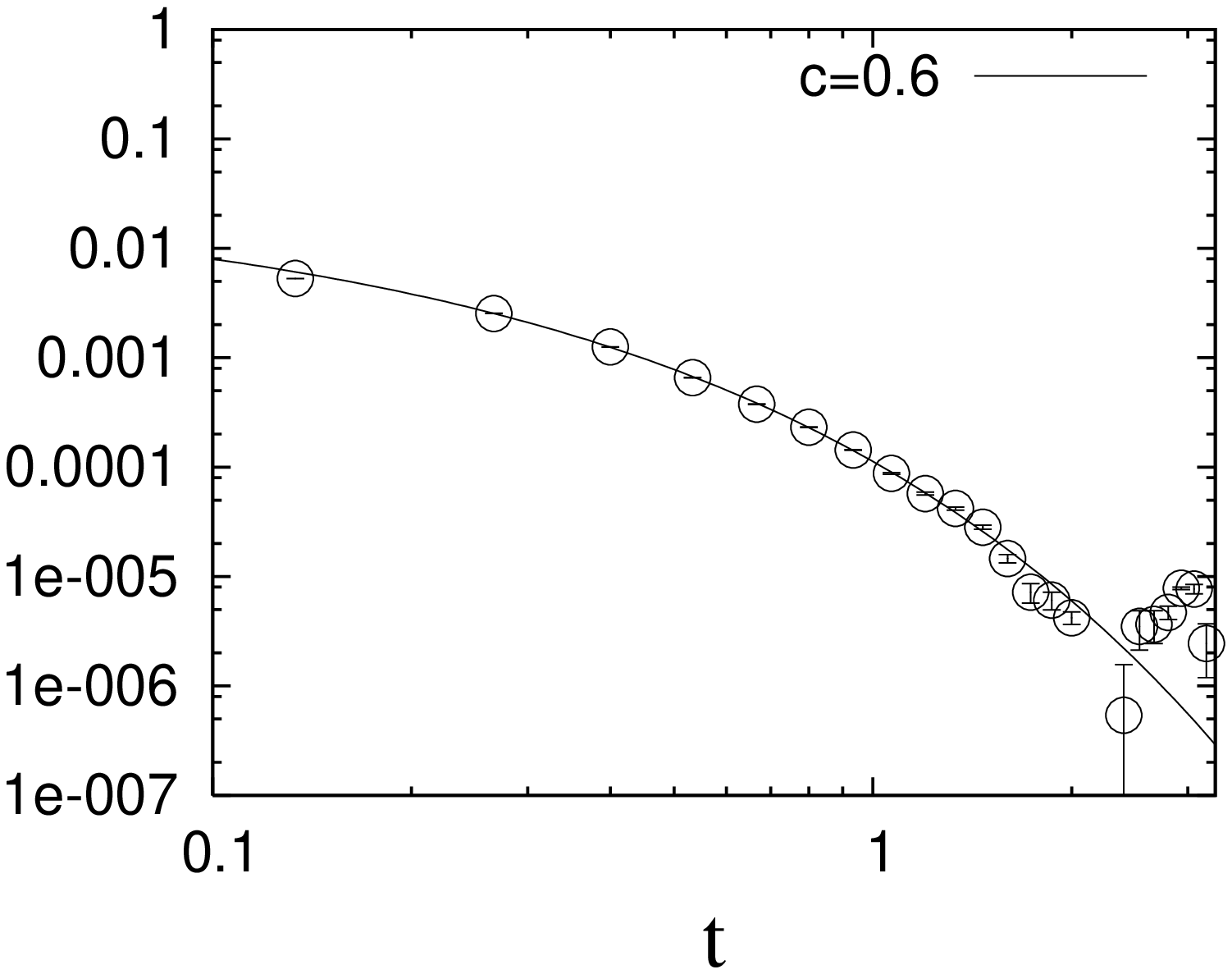,%
angle=270,width=7.4cm}
\caption{(Left) The two-point correlation function of the non-BPS operator
${\cal O}_{kl}$ is plotted for 
$N=3$, $1/\beta=0.15$, $\Lambda=16$.
The curves represent the fits to the form
$a\exp\left(-b \, t^{c}\right)$ with $c=0.3,0.6,0.9$. 
Fitting the data in the region $0.2\le t\le 1.5$ gives 
$c=0.58(2)$.
(Right) The log-log plot of the same correlation function, 
which reveals clear deviation from a straight line behavior
corresponding to the power law.
The solid line represents a fit to the
behavior $a\exp\left(-b \, t^{c}\right)$ with $c=0.6$. 
}
\label{fig:nonSUGRA_T015}
}

Next we consider a non-BPS type operator of the form
\begin{eqnarray}
{\cal O}_{kl}
\equiv
{\rm Tr} 
\left( F_{kl} \, \sum_{i=1}^9 (X_i)^2   \right)
\ , 
\label{nonbmn}
\end{eqnarray} 
where $F_{kl}= -i \, [X_k , X_l]$.
Note that this operator does not belong to the type 
of operators $J^{+}_{\ell}$ in (\ref{currents1})
since the traceless condition is not satisfied.
In fig.~\ref{fig:nonSUGRA_T015}
we show our results for the two-point correlation function 
$\Bigl\langle {\cal O}_{kl}(t) \, {\cal O}_{kl}^{\dagger}(0)\Bigr\rangle$,
which look surprisingly similar to the results for the BMN-type operators.
The log-log plot shown on the right reveals clear deviation from 
a straight line behavior corresponding to the power law.
Fitting the data in the region $0.2\le t\le 1.5$ 
to the form $a\exp\left(-b \, t^{c}\right)$, we obtain $c=0.58(2)$,
which is consistent with $c=3/5$.
This suggests the existence of a universal mechanism for 
generating a mass scale in the case of non-supergravity operators.


%


\section{Summary and discussions}

In this paper we have studied the strong coupling dynamics
of Matrix theory by investigating
the IR behavior of two-point correlation functions.
As we have reviewed in section \ref{sec:gravity_side},
various predictions can be obtained
from the bulk supergravity analysis
based on the gauge-gravity correspondence. 
The direct calculations on the gauge theory side
are completely out of reach of perturbative analyses, however.
This motivated us to perform detailed Monte Carlo evaluation of 
the gauge-theory correlators.
Let us briefly summarize the salient features of our results. 

\begin{enumerate}
\item[{(a)}]The correlation functions of operators corresponding to 
the supergravity modes show the power-law behavior
with the power consistent with the prediction.
Note, in particular, that our Monte Carlo results are obtained
at $N=2$ and $3$, while the supergravity analysis is valid naively
at $N=\infty$.
Finite $N$ effects, if they exist, would be of the order of
$1/N^2$, which is 25\% and 11\% for
$N=2$ and $N=3$, respectively.
Since the agreement with the supergravity predictions is
much more accurate,
it is conceivable that 
the power
is actually independent of $N$ ($\ge 2$).  
\item[{(b)}] The power-law behavior seems to extend
beyond the range (\ref{region}) suggested 
by the validity of the supergravity analysis.
Note that our results for $N=2$ shown 
in fig.~\ref{fig:Jp_N3C16T025} (Left), 
for instance, include the data around $ t \sim 1.5$,
while the upper end of the range (\ref{region})
is around $ t \sim 1.4$ and 1.7 for 
$N=2$ and 3, respectively.
It seems reasonable to consider that the power law would 
continue to be valid beyond this bound 
if we increase the IR cutoff $\beta$.
Indeed, the results in the momentum space shown 
in fig.~\ref{fig:Jp2_MomemtumSpace_mono_v1} suggest
that the power law is valid even at $p/(2\pi) \sim 0.2$,
which is much smaller than 
$1/ t\sim (1.5)^{-1}\sim 0.72$. 
This points to the interesting possibility 
that the power law with the particular power
predicted within the range (\ref{region})
is actually valid even in the far IR region $t\sim N
\rightarrow \infty$ corresponding to M-theory. 
We consider it important
to explore the implication of this statement
on the wave function of the 11-dimensional graviton.  
In particular, the meaning of the anomalous behavior \cite{Y} 
with respect to the 11D boost transformation 
discussed in section \ref{sec:prediction-sugra} must be clarified.
\item[{(c)}] As for the stringy excitations, we only have
semi-quantitative predictions for the BMN-type 
operators with large $J$ and large wave number $n$.
%
Our Monte Carlo results for the BMN-type 
operators with $J=2,3$ provide some evidence
that the predicted form $\exp (-b \, t^{3/5})$ is valid even at small $J$,
although the coefficient $b$ has $J$-dependence 
different from what is expected at large $J$.
The same form is also found to fit well
our results for other non-BPS operators corresponding to 
stringy excited modes.
Our results therefore suggest that 
this peculiar behavior 
may be a universal feature of the mechanism for mass scale 
generation in Matrix theory.
\end{enumerate}

In order to understand the physical meaning of 
the observed behaviors of the correlation functions,
we have discussed the spectral representation of the correlation 
functions in appendix \ref{sec:spectral-rep}.
The power-law behavior for the supergravity modes 
can be naturally understood as a consequence 
of the existence of the zero-energy bound state. 
On the other hand, the exponential behavior 
$\exp (-b \, t^{3/5})$ 
for the stringy excited modes corresponds to 
a very peculiar form of the spectral function.
It would be interesting to clarify possible implications 
of these properties on M-theory.

Let us list some possible directions for future works.
First of all, it is obviously 
worthwhile to extend the Monte Carlo studies
to larger $N$ and larger $\beta$,
which would further clarify our observations (a) and (b) given above. 
It would be also important to study various other operators
corresponding to stringy excited modes and to test
the universal behavior stated in (c).
On the technical side,
it is important to clarify the issue related to the sign problem
in Monte Carlo simulation discussed in appendix \ref{sec:sign-problem}.

From the viewpoint of testing the gauge-gravity correspondence,
it is also interesting to pay attention to 
the normalization, including the sign, of the correlation functions.
For instance, the Ward-like identities for correlation functions 
are discussed \cite{Kanitscheider:2008kd}
using the gauge-gravity correspondence 
in the non-conformal case including the case of D0-branes. 
It would be interesting to test such identities 
by our Monte Carlo method, which would also serve as a
self-consistency check of the method itself. 

From a more general perspective of the gauge-gravity correspondence,
it would be interesting to extend our work
to gauge theories in higher dimensions.
Of particular interest is to 
test the
AdS-CFT correspondence
by studying four-dimensional ${\cal N}=4$ SU($N$)
SYM by Monte Carlo methods,
since most of the tests given so far are not quantitative
except for some special cases in which perturbative analyses 
on the gauge theory side turned out to be useful. 
In fact, a first step in this direction is taken already.
In refs.~\cite{Berenstein_sim}, 
matrix quantum mechanics of 6 bosonic commuting matrices \cite{Berenstein:2005aa}
is shown to give results consistent with
the AdS-CFT correspondence for the three-point functions of chiral primary operators.
More recently,
Monte Carlo studies of the 4d ${\cal N}=4$ SYM
have been performed \cite{Nishimura:2009xm,Honda:2010nx}
based on the novel large-$N$ reduction \cite{Ishii:2008ib},
which reduces the calculation in the SYM (in the planar large-$N$ limit)
to that in the BMN matrix model \cite{BMN}.
The advantage of this method is that supersymmetry
is maximally respected.
In particular, there is no need for fine-tuning the parameters
in the regularized theory unlike the proposals based on the lattice
regularization \cite{latticeSUSY_N4}.
See ref.~\cite{Hanada:2010kt} for a proposal
to combine a two-dimensional lattice with a fuzzy sphere
in order to study the 4d theory at finite $N$ without fine-tuning.

It is also interesting to study
the two-dimensional ${\cal N}=8$ SU($N$)
SYM by Monte Carlo methods.
This corresponds to a system of coincident D1-branes, 
which has an interesting connection 
\cite{Aharony:2004ig} to the black hole/black string transition
as exhibited by the phase diagram with respect to
the temperature and the spatial volume. 
This system has been studied
in the strongly coupled (low temperature) regime
by using the dual type II supergravity, and 
the existence of a first-order transition 
is conjectured \cite{Aharony:2004ig}. 
On the other hand, the weak coupling 
(high temperature) regime can be described effectively by 
a bosonic matrix quantum mechanics, in which
two phase transitions of second and third orders are 
found \cite{Kawahara:2007fn,Mandal:2009vz}.
It is interesting to investigate 
how the two transitions merge into one 
as the coupling constant is increased 
(i.e., as the temperature is decreased)
by direct Monte Carlo methods.
A lattice simulation of two-dimensional SYM 
with 4 supercharges and the SU(2) gauge group was started
a few years ago \cite{Kanamori},
and was extended to SU($N$) with $N=2,3,4,5$ \cite{Hanada:2009hq}.
It was shown that fine-tuning of parameters is not necessary in these cases.
Recently the 16 supercharge case \cite{Catterall:2010fx} 
has been studied.
Let us also recall that the (1+1)D SYM with
16 supercharges is termed ``Matrix String Theory''
as a possible non-perturbative description
of type IIA superstring theory in ten dimensions \cite{Dijkgraaf:1997vv}.
This theory is closely related with 
the doubly-compactified supermembranes 
in 11 dimensions \cite{Sekino:2001ai},
and hence with Matrix theory studied in this paper.
It would be very interesting to explore various 
non-perturbative relations between the two
gauge theories by Monte Carlo methods.

To conclude, with the various new ideas on how to treat 
supersymmetry on a computer,
we consider that interesting non-perturbative physics
in supersymmetric gauge theories,
including those relevant to string theory,
has become accessible by Monte Carlo simulation.
We hope that our work not only provides further motivation
to pursue the Matrix theory conjecture, but also encourages
further development
in Monte Carlo studies of supersymmetric gauge theories.



\acknowledgments
We thank M.~Terashi for his contribution
at the early stage of this work.
We are also grateful to O.~Aharony, Y.~Kikukawa
and K.~Skenderis
for useful discussions and comments. 
Computations have been carried out on PC clusters
at KEK
and Yukawa Institute.
The present work is supported in part by Grant-in-Aid 
for Scientific Research 
(No.\ 20540286 and 23244057 for J.N.,
No.\ 21740216 for Y.S.\
and No.\ 20340048 for T.Y.)
from Japan Society for the Promotion of Science.
The work of M.~H.\ is supported 
from Postdoctoral Fellowship for Research Abroad
by Japan Society for the Promotion of Science.

\appendix



\section{Spectral representation of the correlation functions}
\label{sec:spectral-rep}


In this section we discuss the physical meaning of 
the two-point correlation functions in Matrix theory suggested
from the gauge-gravity duality and from direct Monte Carlo calculation.
For that purpose we consider the spectral representation
\begin{alignat}{3}
\label{spectral-rep}
\Bigl\langle O(t) \, O(0) \Bigr\rangle 
& =
\int_0^{\infty} d\mu \, \rho(\mu) \, \Delta(t, \mu) \ , \\
\Delta (t, \mu) & =
\int \frac{dp}{2\pi}\,  \frac{e^{ipt}}{p^2+\mu^2}
\sim \frac{1}{2\mu}e^{-\mu|t|} 
\label{def-Delta}
\end{alignat}
in the Euclidean space.

Let us first discuss the form (\ref{sugraresult})
 for the supergravity operators.
This corresponds to the spectral representation
$\rho(\mu)\sim \mu^{2\nu+1}$. 
For $2\nu+1>0$, the $\mu$-integral in (\ref{spectral-rep})
is convergent and one retrieves (\ref{sugraresult}). 
For $2\nu+1<0$, we again invoke analytic continuation in 
$\nu$. 
This is in sharp contrast to
the free-theory behavior $|t|^n \, \, 
(n=1, 2, 3, \cdots)$, which corresponds the spectral functions 
with delta-function-like singularities 
at $\mu=0$. 
The milder power-law singularity
in the spectral function 
may be regarded as 
a signature for the existence of 
a zero-energy bound state as a result of very 
non-trivial IR dynamics. The fact that 
$\nu$ increases with the angular momentum $\ell$ 
is consistent qualitatively with the existence of 
many-body states which consist of the zero-energy bound states, 
since the distance 
scale of such states can grow faster 
for larger $\ell$ with a power-law 
behavior as the energy increases.

Let us next discuss the operators
corresponding to stringy excited modes.
According to the scaling argument in sections
\ref{sec:prediction-sugra} and \ref{sec:pred-stringy},
the general form of two-point correlators 
is given by
\begin{equation}
\Bigl\langle O(t) \, O(0) \Bigr\rangle \sim q^{(\Delta +6)/5}
\, \frac{1}{g_s^{\, 2} \, \ell_s^{\, 8}} \, |t|^{-(7\Delta +12)/5}
\, 
f\left( \frac{g_s N|t|^3}{\ell_s^{\,3}}\right) 
\ .
\label{powerlaw}
\end{equation}
For the BMN-type operators with large angular momentum $J$
and 
a large wave number $n$, 
one obtains the prediction
for the function $f(g_sN|t|^3/\ell_s^{\, 3})$ 
at large distances as
\cite{ASY}\footnote{The time coordinate
used in the first paper of ref.~\cite{ASY} is rescaled 
from ours by $t\rightarrow q^{1/2}t$.} 
\begin{equation}
f\left( \frac{g_s N|t|^3}{\ell_s^{\,3}}\right) 
\sim \exp\left(- K \, \frac{q^{1/5}|t|^{3/5}}{J} \right)
\label{massive}
\end{equation}
apart from a possible power-law prefactor,  
with $K$ being an operator-dependent constant.
As the spectral function corresponding to the
behavior (\ref{massive}), let us consider
\begin{equation}
\rho(\mu)\sim \exp \Bigl( C\mu^{a+1} \Bigr) \ ,
\label{rho-ansatz}
\end{equation}
up to a possible power-law correction factor.
At large $|t|$, the $\mu$-integral is dominated by 
the saddle point $\mu=\Bigl( |t|/(a+1)C \Bigr)^{1/a}$.
Comparing the resulting integral with (\ref{massive}), we can determine
the parameters $a$ and $C$.
Plugging them into (\ref{rho-ansatz}), we get
\begin{equation}
\rho(\mu) \sim \exp\left\{-
\frac{2}{3}\left(
\frac{3K}{5 J}
\right)^{5/2}q^{1/2}\mu^{-3/2}\right\} \ .
\label{nonBPSspectralfunction}
\end{equation} 
This is quite different from 
a simple delta-function 
$\rho(\mu) \sim \delta(\mu-M)$,
which represents the standard 
asymptotic behavior of the propagator $e^{-M|t|}$ 
for a massive particle state.
At low energy $\mu \rightarrow 0$,
the spectral function 
(\ref{nonBPSspectralfunction})
decreases exponentially,
%
which suggests that
the {\it single-body} zero-energy 
bound state does not directly contribute to the intermediate states.
On the other hand, the expression (\ref{nonBPSspectralfunction}) 
approaches a constant if we naively extrapolate it to larger $\mu$.
However, it is expected that 
the power-law prefactor in (\ref{powerlaw}) becomes important in this regime.
Since the prefactor is essentially 
the same as in the case of operators corresponding to supergravity modes 
due to the scaling symmetry,
we consider that
the {\it many-body} states consisting 
of the zero-energy bound state play a dominant role 
at high energy.

\section{The algorithm for Monte Carlo simulation}
\label{sec:detail-MC}

In this section we explain the details of the 
algorithm we use for Monte Carlo simulation.
In order to treat fermions efficiently,
it is crucial to use the idea of the 
Hybrid Monte Carlo (HMC) algorithm \cite{Duane:1987de}.

Let us first discuss the idea 
for the bosonic part (\ref{bosonic_action})
neglecting the fermionic part.
We introduce the (fictitious) momentum variables
\beq
\Pi^{ab}_i (t) = \sum_{n=-\Lambda}^{\Lambda} 
\tilde{\Pi} ^{ab} _{i n}
\, e ^{i \omega n t} \ ,
\eeq
which are conjugate to $X^{ab}_i(t)$.
Due to the Hermiticity of $\Pi^{ab}_i (t)$,
we have the constraint 
\beq
\tilde{\Pi}^{ab}_{i, -n} = (\tilde{\Pi}^{ba}_{i n})^{*} \ .
\eeq
Similarly, we introduce the momentum variables $p_a$ corresponding
to $\alpha_a$.
Then we define the fictitious Hamiltonian for the bosonic part as
\beq
H_{\rm b}= \frac{1}{2} \sum_{a=1}^{N} (p_a)^2
+ \frac{1}{2} 
\sum_{n=-\Lambda}^{\Lambda} 
\tilde{\Pi}^{ab}_{i n} \, \tilde{\Pi}^{ba}_{i,-n}
+ S_{\rm b}[X,\alpha] \ .
\label{H-boson}
\eeq
The Hamilton equation can be obtained as
\beqa
\frac{d \tilde{X}^{ab}_{i n}}{d \tau}
&=& \frac{\del H_{\rm b}}{\del \tilde{\Pi}^{ab}_{i n}}
= \tilde{\Pi}^{ba}_{i, - n} \\
\frac{d \tilde{\Pi}^{ab}_{i n}}{d \tau}
&=& - \frac{\del H_{\rm b}}{\del \tilde{X}^{ab}_{i n}}
= - \frac{\del S_{\rm b}[X,\alpha]}{\del \tilde{X}^{ab}_{i n}} \ .
\label{Pi-evolve}
\eeqa
By using the explicit form of $S_{\rm b}[X,\alpha]$,
we can calculate the second line explicitly as
\beqa
\frac{d \tilde{\Pi}^{ab}_{i n}}{d \tau}
&=& - N \beta \Bigl[ 
\Bigl( n \omega - \frac{1}{\beta}(\alpha_a - \alpha_b) \Bigr)^2
\tilde{X}^{ba}_{i, -n}  \nonumber \\
&~& \quad \quad 
+  2 (\tilde{X}_{i}\tilde{X}_{j}\tilde{X}_{j} )^{ba}_{-n} 
 -  4 (\tilde{X}_{j}\tilde{X}_{i}\tilde{X}_{j} )^{ba}_{-n}
+  2 (\tilde{X}_{j}\tilde{X}_{j}\tilde{X}_{i} )^{ba}_{-n} 
\Bigr] \ .
\eeqa
Similarly, the Hamilton equations for the gauge variables
are obtained as
\beqa
\frac{d \alpha_a}{d \tau}
&=& \frac{\del H_{\rm b}}{\del p_a} = p_a \\
\frac{d p_a}{d \tau}
&=& - \frac{\del H_{\rm b}}{\del \alpha_a}
= 2 N 
\sum_{n=-\Lambda}^{\Lambda} 
\Bigl( n \omega - \frac{1}{\beta}(\alpha_a - \alpha_b) \Bigr)
\tilde{X}^{ab}_{i n} \, \tilde{X}^{ba}_{i, -n} \ .
\label{p-evolve}
\eeqa
Updating the bosonic variables is performed by
(i) refreshing the momentum variables 
$p_a$ and $\tilde{\Pi} ^{ab} _{i n}$ by Gaussian random numbers,
which obey the distribution $\ee^{-H_{\rm b}}$
and 
(ii) solving the Hamilton equations for a fixed time interval $\tau$.
In order to satisfy the detailed balance,
one needs to adopt the so-called leap-frog discretization for 
$\tau$-evolution, and to perform the Metropolis accept/reject procedure
after (ii) with the acceptance probability 
$\min (1, e^{-\Delta H_{\rm b}})$.
For more details, see refs.~\cite{Ambjorn:2000bf,Ambjorn:2000dx},
in which the method is described for supersymmetric matrix models.

%


Let us next introduce fermions. 
We expand the Fourier component $\tilde{\psi}_{\alpha n}$
of the fermion field as
\beq
\tilde{\psi}_{\alpha n}
= \sum_{A=1}^{N^2} \tilde{\psi}_{\alpha n}^A \, t^A
\eeq
in terms of U($N$) generators $t^A$.
(We will impose the traceless condition later.)
Here we use a representation given by
\beq
(t^A)_{ab} = \delta_{a i_A} \delta_{b j_A} \ , 
\eeq
where $a=1, \cdots , N^2$, and we have defined
\beqa
A &=& N(i_A - 1)+ j_A \ , \\
\bar{A} &=& N(j_A - 1)+ i_A \ ,
\eeqa
with which we may write
\beq
\Tr 
(t^A t^B) = \delta_{\bar{A} B} \ .
\eeq
Let us also define
\beq
g_{ABC} \equiv 
\Tr
\Bigl( t^C [t^A , t^B] \Bigr)
= \delta_{j_A i_B} \delta_{j_B i_C} \delta_{j_C i_A}
- \delta_{j_C i_B} \delta_{j_B i_A} \delta_{j_A i_C} \ .
\eeq
Then the fermionic action $S_{\rm f}$ 
given in eq.~(\ref{bfss_action_cutoff}) may be
written in the form
\beq
S_{\rm f}
= \frac{1}{2} \, {\cal M} ' _{A \alpha n ; B \beta p}
\, \tilde{\psi}_{\alpha n}^A \, \tilde{\psi}_{\beta p}^B \ ,
\eeq
where we have defined a matrix ${\cal M} '$ 
of dimension $16(2\Lambda+1)N^2$ by
\beq
\mathcal{M}_{A \alpha p, B \beta q} '
= - \beta \, (\gamma_i)^{\alpha \beta} \, 
\tilde{X}_{i p-q}^C \, g_{ABC}
+ \beta \, \delta_{pq} \, 
\Bigl( i \, p \, \omega - i \frac{1}{\beta} 
(\alpha_{i_B} - \alpha_{j_B} ) \Bigr) \, 
\delta_{\alpha\beta}  \, 
\delta_{\bar{A}B} \ .
\eeq
We impose the traceless condition on the fermion field
by making the replacement
\beq
(\tilde{\psi}_{\alpha n})^{NN} \mapsto
- \sum_{a=1}^{N-1} (\tilde{\psi}_{\alpha n})^{aa} \ .
\eeq
The fermion action can be written in terms of the
remaining degrees of freedom as
\beq
\tilde{\psi}^A_{\alpha p} \, 
\mathcal{M}' _{A \alpha p , B \beta q} \, 
\tilde{\psi}^B_{\beta q}
= \tilde{\psi}^{A'}_{\alpha p} \, 
\mathcal{M}  _{A' \alpha p , B' \beta q} \, 
\tilde{\psi}^{B'}_{\beta q} \ ,
\eeq
where we have defined 
a matrix ${\cal M} $ 
of dimension $16(2\Lambda+1)(N^2-1)$ by
\beqa
\mathcal{M}  _{A' \alpha p , B' \beta q}
&=& \mathcal{M} ' _{A' \alpha p , B' \beta q}
- \mathcal{M} ' _{N^2 \alpha p , B \beta q} \delta_{i_{A'} j_{{A'}}} 
\nonumber \\
&~& - \mathcal{M} ' _{A ' \alpha p , N^2  \beta q} \delta_{i_{B'} j_{{B'}}}
+ \mathcal{M} ' _{N^2 \alpha p , N^2 \beta q}
\delta_{i_{A'} j_{{A'}}} \delta_{i_{B'} j_{{B'}}} \ .
\label{def-calM}
\eeqa
Integrating out the fermions,
we obtain
${\rm Pf}{\cal M}$, which is complex
in general. As we explained below (\ref{bfss_action_cutoff}),
we simply replace it by its absolute value
\beq
\Bigl|{\rm Pf}{\cal M} \Bigr|
= {\rm det} \, \Bigl( {\cal D}^{1/4} \Bigr) \ ,
\eeq
where ${\cal D}={\cal M}^\dag {\cal M}$.

The trick of the RHMC algorithm \cite{Clark:2003na} is to
represent $|{\rm Pf}{\cal M}|$ 
as
\beq
\Bigl|{\rm Pf}{\cal M} \Bigr|
= \int dF dF^*  e ^{-S_{\rm PF}} \ ,
\eeq
where
\beq
S_{\rm PF} =
a_0 \, F^*  F + \sum_{k=1}^{Q}
a_k \, F^* ({\cal D}+b_k)^{-1} F \ ,
\label{PF-pf}
\eeq
%
using the auxiliary complex variables $F$,
which are called pseudo-fermions.
Here the constants $a_k$ and $b_k$ are real positive parameters
appearing in the rational approximation
\beq
x^{-1/4} \simeq
a_0 + \sum_{k=1}^{Q}
\frac{a_k}{x+b_k}  \ ,
\label{rational-approx}
\eeq
and they can be generated by a code \cite{Clark-Kennedy} based on 
the Remez algorithm.
The approximation (\ref{rational-approx}) can be made to
have sufficiently small relative errors (smaller than $\delta$)
within a certain range ($\epsilon < x < 1$).
%
%
%
In our simulation, we use $Q=15$ with $\delta=1.19\times 10^{-4}$ and
$\epsilon=10^{-12}$.
We rescale the matrix ${\cal D}$ by an appropriate
constant factor so that the largest eigenvalue of ${\cal D}$ is
well below the upper bound for the approximation.

Then we apply the usual
HMC algorithm to the whole system.
The Hamiltonian for the pseudo-fermions is given by
\beq
H_{\rm PF} = \sum_I \Phi_I ^{*} \, \Phi_I + S_{\rm PF} \ ,
\eeq
where the field $\Phi$ represents the momentum variables 
conjugate to the pseudo-fermions $F$.
Here and henceforth, the index $I$ is used to represent 
the spinor index $\alpha$, the momentum index $n$
and the SU($N$) index $A$, collectively.
The Hamilton equations for the pseudo-fermions are given by
\beqa
\frac{d F_{I}}{d \tau}
&=& \frac{\del H_{\rm PF}}{\del \Phi_{I}}
= \Phi_{I}^{*} \\
\frac{d \Phi_{I}}{d \tau}
&=& - \frac{\del H_{\rm PF}}{\del F_{I}}
= - a_0 \, F_I ^{*} - 
\sum_{k=1}^Q a_k \, G_I^{(k)*} \ ,
\eeqa
where $G_I^{(k)}$ is defined by
\beq
\Bigl(\mathcal{M}^\dag \mathcal{M} + b_k \Bigr)_{IJ} \, G_J^{(k)} = F_I 
\quad \quad \mbox{for $k=1,\cdots ,Q$}  \ .
\label{lin-eq-bfss}
\eeq
We have to add extra terms
\beqa
 - \frac{\del S_{\rm PF}}{\del \tilde{X}^{ab}_{i n}} 
&=& \sum_{k=1}^{Q} a_k \, G^{(k)*} 
\left( 
\mathcal{M}^\dag \, \frac{\del \mathcal{M}}{\del \tilde{X}^{ab}_{i n}}
+ \frac{\del \mathcal{M}^\dag }{\del \tilde{X}^{ab}_{i n}} 
\, \mathcal{M} \right)  G^{(k)}  \ , \\
 - \frac{\del S_{\rm PF}}{\del \alpha_a}
&=& \sum_{k=1}^Q a_k \,
G^{(k)*} 
\left( \mathcal{M}^\dag \, \frac{\del \mathcal{M}}{\del \alpha_a}
+ \frac{\del \mathcal{M}^\dag }{\del \alpha_a} 
\, \mathcal{M} \right) G^{(k)} 
\eeqa
on the right-hand side of
eqs.\ (\ref{Pi-evolve}) and (\ref{p-evolve}), respectively.

The main part of the computation
comes from
solving a linear system (\ref{lin-eq-bfss}).
We solve the system
for the smallest $b_k$
using the conjugate gradient
method, which reduces the problem to
the iterative multiplications of ${\cal M}$
to a pseudo-fermion field,
each of which requires O($\Lambda^2 N^3$)
arithmetic operations.\footnote{Note that one should not construct
${\cal M}$ explicitly and multiply it to 
a pseudo-fermion field literally,
which would require ${\rm O}(\Lambda^2 N^4)$ arithmetic operations.
Instead one should reduce the procedure to 
multiplication of $N \times N$ matrices using the original
definition of ${\cal M}$.}
%
The solutions for larger $b_k$'s 
can be obtained as by-products
using the idea of the multi-mass 
Krylov solver \cite{Jegerlehner:1996pm}. 
This avoids the factor
of $Q$ increase of the computational effort.
When we solve the fictitious classical
Hamilton dynamics, the step size of the discretized evolution 
may depend on the Fourier mode.
We take the step size to be proportional to the average
fluctuation of each mode of the bosonic matrices
so that
the configuration space can be swept out 
most efficiently. 
This technique is called the Fourier acceleration 
\cite{Catterall:2001jg}. In the lattice gauge theory, 
in order to apply the Fourier acceleration one has to transform the 
configuration to the momentum representation. 
On the other hand, in the present momentum cutoff method, 
the Fourier acceleration can be implemented without any additional cost
since we are working directly in the momentum space.
Thanks to this advantage, the efficiency of the algorithm
is enhanced drastically.
%

In actual simulation, we observe certain 
instability \cite{AHanada-Nishimura-Takeuchi},
which is related to the existence of the flat direction
$[X_i , X_j] \approx 0$ in the potential term of the action (\ref{cQM}).
This instability can be seen by probing the observable
\beq
R^2 \equiv  \frac{1}{N\beta} \int_0^\beta dt \,
\Tr 
(X_i)^2 \ .
\label{defR}
\eeq
Typically this quantity fluctuates around some value,
but sometimes it grows rapidly and the simulation bumps.
In order to avoid this problem, we introduce a cutoff on $R^2$,
which is taken to be sufficiently 
larger than the upper edge of the fluctuation.
The value of the cutoff is $5.0$ and $3.8$ for $N=2$ and $N=3$, respectively. 
%

\section{Sign problem}
\label{sec:sign-problem}

In this section we discuss the so-called sign problem
in Monte Carlo studies of Matrix theory (\ref{cQM}).
As we explained below eq.~(\ref{bfss_action_cutoff}),
we neglect the phase of the Pfaffian that appears
from integrating out fermionic matrices.
%



    \FIGURE[t]{
    \epsfig{file=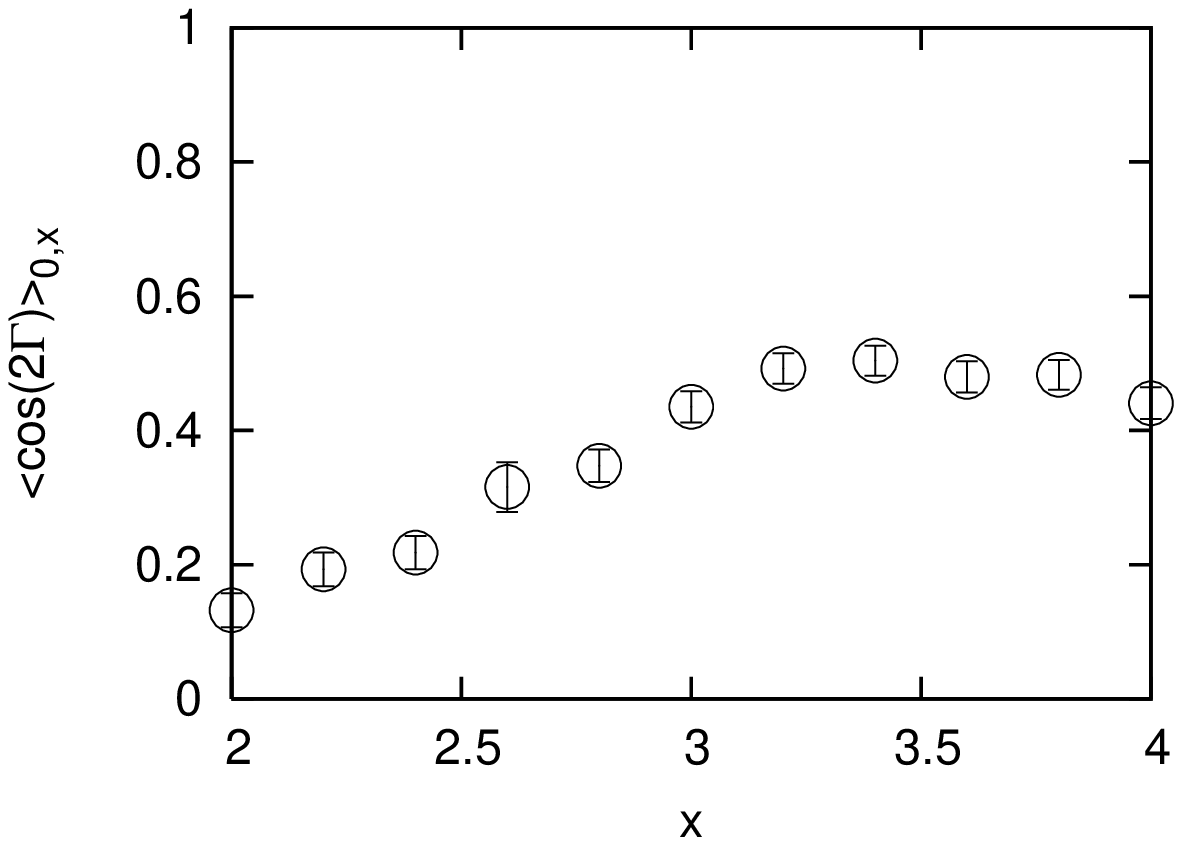,%
angle=270,width=7.4cm}
    \epsfig{file=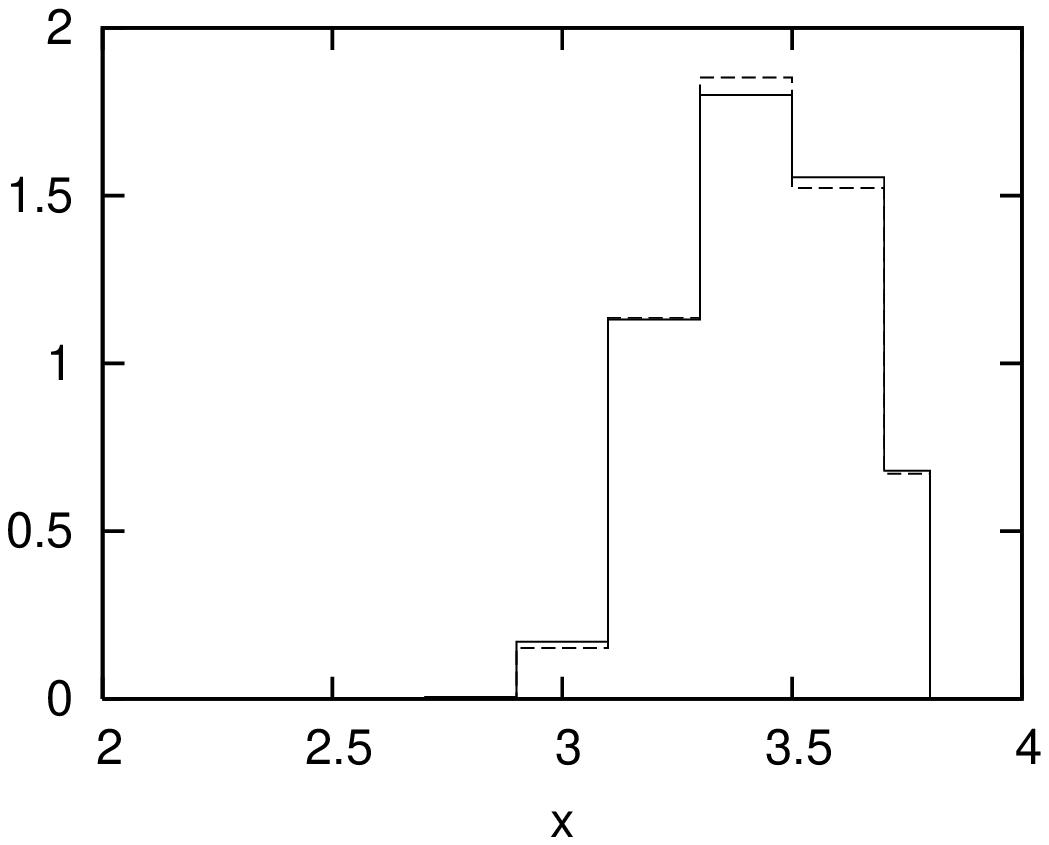,%
angle=270,width=7.4cm}
\caption{(Left) The VEV $\langle\cos 2\Gamma\rangle_{0,x}$ is plotted for
SU(3), $\beta = 6.67$, $\Lambda=8$. 
(Right) The distribution $\rho^{(0)}(x)$ 
and the product 
$\tilde{\rho}(x) \propto \rho^{(0)} (x) 
\tilde{w}(x)$
with the normalization $\int dx  \tilde{\rho}(x) =1$
are plotted by the solid line and the dashed line, respectively.
SU(3), $\beta = 6.67$, $\Lambda=8$. 
}
\label{fig:TrX2}
}

Here we present some numerical evidence, which shows that 
the effect of the phase is indeed small.
As a typical observable, let us consider $R^2$ defined by
(\ref{defR}),
and discuss how it is 
affected by the fluctuation of the phase.
%
For that purpose we use 
the factorization method \cite{Anagnostopoulos:2001yb,%
Ambjorn:2002pz,Anagnostopoulos:2010ux},
which was proposed to study the effect of the phase in a general
model suffering from the sign problem.
Let us define the distribution 
function
\beqa
\rho (x) &=& \Bigl\langle \delta (x-R^2) \Bigr\rangle   \ , \\
\rho^{(0)} (x) &=& \Bigl\langle \delta (x-R^2) \Bigr\rangle_0 \ ,
\eeqa
for the full model and for the phase-quenched model, respectively.
One can then easily show that 
\beq
\rho(x) \propto \rho^{(0)} (x) \, w(x) \ ,
\eeq
where the correction factor $w(x)$ is given by
\beq
w(x) \equiv \Bigl\langle e^{i \Gamma} \Bigr\rangle_{0,x} 
=  \Bigl\langle \cos \Gamma \Bigr\rangle_{0,x} \ .
\eeq
Here, $\Gamma$ represents the phase of the Pfaffian,
and the symbol $\langle \ \cdot \  \rangle_{0,x}$ represents
a VEV with respect to the phase-quenched model with the constraint
$R^2 = x$.

Since the calculation of 
${\rm Pf} {\cal M}$
is time-consuming,
we calculate 
${\rm det} {\cal M} = ({\rm Pf} {\cal M})^2$,
from which we can obtain
$\tilde{w}(x) \equiv \Bigl\langle \cos (2 \Gamma) \Bigr\rangle_{0,x}$.
This is sufficient for estimating an upper bound on the effect
of the phase
since the factor of 2 in the cosine only magnifies it.
In fig.~\ref{fig:TrX2} (Left) we plot 
$\tilde{w}(x)$ for $N=3$, $\beta=6.67$ and $\Lambda=8$.
In fig.~\ref{fig:TrX2} (Right) we plot the distribution function
$\rho^{(0)} (x)$ for the phase-quenched model
and the product 
$\tilde{\rho}(x) \propto \rho^{(0)} (x) \, \tilde{w}(x)$
with the normalization $\int dx \,  \tilde{\rho}(x) =1$.
The difference between
$\tilde{\rho}(x)$ and $\rho^{(0)} (x)$
is indeed negligible, which implies that
the expectation value $\langle R^2 \rangle$ is 
not affected by the effect of the phase.

We speculate that the phase quenching can be completely justified
in the large-$\beta$ limit. 
At large $\beta$,
the expectation value of $\langle R^2 \rangle$
can be obtained by solving the saddle-point equation
\beq
\frac{d}{dx} \ln \rho^{(0)}(x) = - \frac{d}{dx} \ln w(x) \ .
\label{master-eq}
\eeq
First, we can easily prove that 
the Pfaffian becomes real
if we omit the kinetic term in (\ref{bfss_action_cutoff}).
As $\beta$ increases, 
there are actually more and more low-momentum modes,
for which the kinetic term is small.
Therefore,
we consider it conceivable that the fluctuation of the phase
does not increase as ${\rm O}(\sqrt{\beta})$, which
is the typical growth of the fluctuation for extensive quantities
at large $\beta$.
As a result, the right-hand side of (\ref{master-eq})
is expected to be ${\rm O}(\beta^p)$ with $p<1$,
whereas the left-hand side 
is expected to be ${\rm O}(\beta)$
according to the usual scaling argument.
In fig.~\ref{fig:TrX2}
we do observe that the $x$ dependence of 
$\tilde{w}(x)$ is much smaller than that of $\rho^{(0)} (x)$.
If our scenario is correct,
we can neglect the effect of the phase completely
in the $\beta \rightarrow \infty$ limit.
In order to confirm this scenario, we need to investigate
the $\beta$ dependence, which we leave for future investigations.
%


\end{document}